\documentclass{jfm}

\usepackage{graphicx}
\usepackage{newtxtext}
\usepackage{newtxmath}
\usepackage{natbib}
\usepackage{hyperref}
\hypersetup{
    colorlinks = true,
    urlcolor   = blue,
    citecolor  = black,
}

\newcommand{\RomanNumeralCaps}[1]
\linenumbers

\usepackage{graphics}
\usepackage{graphicx}
\usepackage{amsmath}
\usepackage{amssymb}
\usepackage{xcolor}
\usepackage{tikz}
\usepackage{siunitx}
\usepackage{booktabs} 
\usepackage{multirow} 
\usepackage{bigdelim}
\usepackage{amssymb,pifont}
\usepackage{derivative}
\usepackage[justification=justified,width=\textwidth]{caption}
\usepackage[T1]{fontenc}

\definecolor{nltw_blue}{rgb}{0,0.6,1}
\definecolor{sw_green}{rgb}{0,0.5647,0}
\definecolor{hj_orange}{rgb}{0.8118,0.3412,0.2745}

\definecolor{color_1}{rgb}{0.9451,0.5569,0.1098}
\definecolor{color_2}{rgb}{0.8824,0.0980,0}
\definecolor{color_3}{rgb}{0,0.0745,0.7490}
\definecolor{color_4}{rgb}{0.7490,0,0.3765}
\definecolor{color_5}{rgb}{0,0.6000,1.0000}
\definecolor{color_6}{rgb}{0.5490,0.7333,0.1490}
\definecolor{color_7}{rgb}{0.0980,0.5020,0}
\definecolor{color_8}{rgb}{0.25,0.25,0.25}

\definecolor{color_9}{rgb}{0.6350,0.0780,0.1840}
\definecolor{color_10}{rgb}{0.0980,0.0980,0.4392}

\DeclareRobustCommand\fullblack {\tikz[baseline=-0.6ex ]\draw[thick] (0,0)--(0.4,0);}
\DeclareRobustCommand\dashedblack {\tikz[baseline=-0.6ex ]\draw[thick,dashed] (0,0)--(0.4,0);}
\DeclareRobustCommand\dasheddottedblack
{\tikz[baseline=-0.6ex]\draw [thick,dash pattern={on 4pt off 3pt on 1pt off 3pt on 4pt}] (0,0) -- (0.54,0);}

\newcommand{\WS}[1]{#1}

\usepackage{subfigure}

\definecolor{color_pwd_1}{HTML}{FF0000}
\definecolor{color_pwd_2}{HTML}{0072BD}
\definecolor{color_pwd_3}{HTML}{000000}
\definecolor{color_pwd_4}{HTML}{007F00}
\definecolor{color_pwd_5}{HTML}{007F00}
\definecolor{color_pwd_6}{HTML}{FFD700}

\definecolor{color_curve_1}{HTML}{A1132E}
\definecolor{color_data_synolakis}{HTML}{B4B4B4}
\definecolor{color_curve_xp}{HTML}{009000}
\definecolor{color_curve_vp}{HTML}{cf5746}
\definecolor{color_curve_noda}{HTML}{7d7d7d}

\newcommand{\Lp}{$\Lambda_p$}
\newcommand{\Fp}{$\mathrm{Fr}_p$}

\title{Nascent water waves induced by the impulsive motion of a solid wall}

\author{Wladimir Sarlin\aff{1,2}
  \corresp{\email{wladimirsarlin@outlook.com}},
  Zhaodong Niu\aff{1,2},
  Alban Sauret\aff{3,4},
  Philippe Gondret\aff{1},
  \and Cyprien Morize\aff{1}}

\affiliation{\aff{1} Universit\'e Paris-Saclay, CNRS, Laboratoire FAST, F-91405 Orsay, France
\aff{2} Laboratoire d'Hydrodynamique, CNRS, École polytechnique, Institut 
Polytechnique de Paris, 91120 Palaiseau, France
\aff{3} University of California, Santa Barbara, Department of Mechanical Engineering, CA 93106, USA
\aff{4} Department of Mechanical Engineering, University of Maryland, College Park, Maryland 20742, USA}

\begin{document}
\maketitle

\begin{abstract}
The description of the generation mechanism of impulse surface waves remains an important challenge in environmental fluid mechanics, owing to the need for a better understanding of large-scale phenomena such as landslide-generated tsunamis. In the present study, we investigated the generation phase of laboratory-scale water waves induced by the impulsive motion of a rigid piston, whose \WS{maximum} velocity \WS{$U$} and \WS{total} stroke \WS{$L$} are independently varied, as well as the initial liquid depth \WS{$h$}. By doing so, \WS{the influence} of two dimensionless numbers \WS{is} studied: the Froude number \Fp \WS{= $U/(gh)^{1/2}$}, with $g$ the gravitational acceleration, and the relative stroke $\Lambda_p =L/h$ of the piston. During the constant acceleration phase of the vertical wall, a transient water bump forms and remains localised in the vicinity of the piston, for all investigated parameters. Experiments with a small relative acceleration $\gamma/g$\WS{, where $\gamma=U^2/L$,} are well captured by a first-order potential flow theory established by \citet{1990_joo}, \WS{which} provides a fair estimate of the overall free surface elevation and the maximum wave amplitude reached at the contact with the piston. For large Froude numbers, however, wave breaking hinders the use of such an approach. \WS{In this case}, an unsteady hydraulic jump theory is proposed, which accurately predicts the time evolution of the wave amplitude at the contact with the piston throughout the generation phase. At the end of the formation process, the dimensionless volume of the bump evolves linearly with $\Lambda_p$ and the wave aspect ratio is found to be governed, at first-order, by the relative acceleration $\gamma/g$. As the piston begins its constant deceleration, the water bump evolves into a propagating wave and several regimes such as dispersive, solitary-like and bore waves, as well as water jets are then reported and mapped in a phase diagram in the (\Fp, $\Lambda_p$) plane. While the transition from waves to water jets is observed if the typical acceleration of the piston is close enough to the gravitational acceleration $g$, the wave regimes are found to be mainly selected by the relative piston stroke $\Lambda_p$. On the other hand, the Froude number determines whether the generated wave breaks or not. 
\end{abstract}


{\bf MSC Codes }  {\it(Optional)} Please enter your MSC Codes here

\section{Introduction}
\label{Sec1}

\WS{The generation of waves at the surface of a liquid is a crucial physical problem for understanding a wide range of phenomena that occur in Nature or in industrial processes, such as periodic water waves generated by the wind at the ocean surface \citep{2019_perrard}, slamming in the context of ship hydrodynamics \citep{2018_dias}, sloshing in moving containers \citep{1984_chwang,2005_ibrahim}, impulse waves generated by landslides \citep{2004_fritz,2021b_robbe-saule,2022_rauter,2024_darvenne} or iceberg calving \citep{2021_wolper}.}

The description of the free surface elevation resulting from an initial impulsion is a long-standing problem in fluid mechanics that can be traced back to the classical works of Cauchy and Poisson at the beginning of the nineteenth-century \citep{1818_poisson,1827_cauchy,2003_darrigol}. Later, surface waves produced by a harmonic forcing have been studied analytically by \citet{1929_havelock}, who employed linear theory to this end [see also \citet{1951a_biesel,1951b_biesel,1951c_biesel} for more detail]. After these seminal contributions, and motivated by the growing number of potential applications of the subject, many studies followed, that investigated waves generated by the impulsive motion of a rigid body, with the aim of describing the free surface elevation in the vicinity of the forcing region. In particular, based on the prior theoretical work made by \citet{1949_kennard}, \citet{1970_noda} derived solutions to the linearized equations for gravity surface waves corresponding to two idealized cases of landslides: the vertical fall of a solid block and the horizontal translation of a rigid wall. In the second scenario, the linear relationship 

\begin{equation}
\frac{A_{m0}}{h} \simeq 1.2\, \mathrm{Fr}_p
\label{NodaEq}
\end{equation}

\noindent was established between the maximum wave amplitude $A_{m0}$ at the contact with the translating piston, the initial fluid depth $h$ and the Froude number $\mathrm{Fr}_p=U/\sqrt{gh}$. This dimensionless quantity compares the forcing velocity $U$ \WS{of the advancing wall} to the celerity $\sqrt{gh}$ of linear gravity waves in shallow water, where $g$ stands for the gravitational acceleration. The linear relationship obtained by \citet{1970_noda} was successfully compared with previous experiments performed by \citet{1966_miller}, and further confirmed by another experimental investigation conducted by \cite{1972_das}. In this study, these authors reported a phase diagram in the (\Fp, $\Lambda_p$) plane, with $\Lambda_p=L/h$ being the ratio between the \WS{total} piston stroke $L$ and the initial water depth $h$, \WS{and \Fp\ being calculated using the average velocity of the advancing wall during the generation phase}. Several wave regimes were identified: two dominated by dispersive effects (the so-called oscillatory and non-linear transition regions), and two revealing an increasing influence of non-linear effects (solitary and bore waves). \WS{Other experimental measurements and theory for the force developing on an accelerating piston in a fluid channel and the free surface elevation have also been reported by \citet{1986_synolakis,1989_synolakis} for different kinds of piston motion. In particular, these studies revealed that the maximum relative amplitude of the wave is very close to the linear law $A_{m0}/h = \mathrm{Fr}_p$ when $\mathrm{Fr}_p < 1$. However, when the Froude number becomes large, two nonlinear theoretical relations are put forward by \citet{1986_synolakis,1989_synolakis} to connect the wave amplitude at the contact with the piston to $\mathrm{Fr}_p$: under shallow water conditions and using the method of characteristics, one gets}
 
\WS{
\begin{equation}
\frac{A_{m0}}{h} = \mathrm{Fr}_p + \frac{1}{4} {\mathrm{Fr}_p}^2
\label{PlateLawEq}
\end{equation}}

\noindent \WS{while, when the plate is generating bores,}

\WS{\begin{equation}
\mathrm{Fr}_p = \frac{A_{m0}}{h} \left(\frac{1 + A_{m0}/\left(2h\right)}{1 + A_{m0}/h}\right)^{1/2}
\label{HJEq}
\end{equation}}

\noindent \WS{results from the mass and momentum conservation equations \citep{1999_whitham}. This second nonlinear law has also been shown to give good predictions for bore waves generated by the collapse of a granular column into shallow water, where the granular front acts like a rigid piston \citep{2021b_sarlin}.
}

In another approach of the problem, several authors such as \citet{1983_chwang}, \citet{1984_lin}, or \citet{1984_chwang} considered the Euler equations and developed methods based on the potential flow assumption and on small-time expansions to model the early generation phase of impulse waves. They derived first and second-order solutions for the free surface elevation and identified a singular behaviour at the contact point with the advancing rigid wall. In particular, \citet{1984_chwang} described the structure of the nascent wave in the case of an impulsive sloshing motion, with accelerated rectangular and cylindrical containers \WS{partially} filled with water. The non-uniformity of the solution was successfully analyzed by \cite{1987_roberts} who developed a theory based on small-amplitude expansions, which was shown to circumvent the singular behaviour at the contact point between the free surface and the solid piston. This aspect was also studied by \citet{1994_king}, who considered the case of waves generated by a uniformly accelerated plate and employed matched asymptotic small-time expansions, which allowed these authors to develop a temporally uniform solution. Similar approaches were followed later by \citet{2007_needham} and \citet{2015_uddin}, who considered the free surface elevation caused by a rigid wall advancing at constant velocity and the influence of weak surface tension effects on the problem, respectively. In the latter case, \citet{2015_uddin} demonstrated that four asymptotic regions have to be studied to correctly describe the induced wave. They successfully compared their analytical solution to experimental measurements of the free surface elevation during the early times of the generation process. Following \cite{1987_roberts}, \citet{1990_joo} developed a theory based on a Fourier integral method and a small Froude number expansion of a potential flow. Their analysis included surface tension and wettability effects and led to the derivation of leading-order solutions for the free surface elevation in various forcing cases including ramp, step or even harmonic velocities imposed at the advancing wall. The asymptotic behaviour of these expressions was thoroughly discussed by these authors, as well as the influence of surface tension which has, for instance, the effect of removing the small wiggles that are observed otherwise in the vicinity of the wavemaker. In a different approach, to obtain a given long wave, it is also possible to solve an inverse evolution problem, as shown by the pioneering works of \citet{1978_goring}, \cite{1980_goring}, and \cite{1990_synolakis}. By applying this method to solutions to the Korteweg--de Vries equation, these authors were able to determine the correct trajectory to confer to a piston wavemaker to produce a given solitary or cnoidal wave. More recently, studies made, amongst others, by \cite{2002_guizien} or \citet{2020_francis} refined this approach to accurately generate solitary waves either in laboratory experiments or numerically.

Despite this extensive research, open questions remain about the generation of impulse surface waves due to the intrinsic complexity of the problem. In particular, the phase diagram of the possible wave regimes provided by \cite{1972_das} is, according to these authors, incomplete, as in their study a weak coupling existed between the stroke and the velocity of the piston. A similar pairing is observed in other configurations, for instance in model experiments studying water waves generated by the gravity-driven fall of a granular medium \citep{2022a_sarlin}. Thus, what happens when such a coupling is removed remains unclear. Another important aspect is to determine to which extent the theoretical models existing in the literature are able to describe waves generated experimentally and give a relevant prediction for the free surface elevation \WS{for the different wave regimes}. 
In this study, we report extensive experiments on impulse surface waves generated by the horizontal translation of a rigid vertical wall in a water flume. In section \ref{sec_apparatus}, the experimental methods, parameters of interest, and the forcing mechanism are presented. This is followed by an analysis of the wave regimes obtained in section \ref{sec_wave_regimes}, which leads to their mapping in a phase diagram. Section \ref{sec_generation} provides a qualitative and quantitative description of the generation phase for different representative examples of impulse waves observed in the experiments. This preludes to a thorough discussion given in section \ref{sec_model} on the manner to predict the transient shape of the induced waves and the maximum amplitude reached at the contact with the moving wall when the generation process ends. Concluding remarks and perspectives for future work are finally given in section \ref{sec_conclusion}.

\section{Experimental apparatus and protocol}
\label{sec_apparatus}

\begin{figure}
	\centerline{\includegraphics[width=\linewidth]{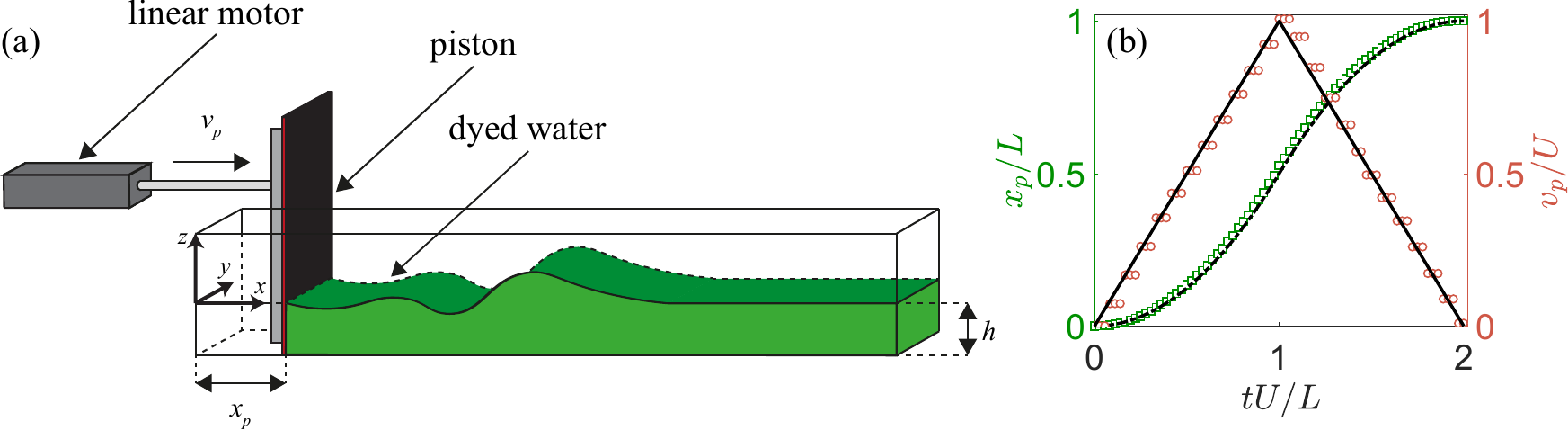}}
	\caption{(a) Schematic view of the experimental setup. (b) \WS{Prescribed} evolution of the relative stroke $x_p/L$ (dash-dotted line) and velocity $v_p/U$ (solid line) of the piston with the dimensionless time $tU/L$, which consists of a constant acceleration phase for a time $0 \leqslant t \leqslant L/U$ followed by a constant deceleration for $L/U \leqslant t \leqslant 2L/U$. \WS{The measured values for the relative stroke (\textcolor{color_curve_xp}{$\square$}) and velocity (\textcolor{color_curve_vp}{$\circ$}) of the piston are also reported for a typical experiment where $L=14.5\ \mathrm{cm}$, $U=1.19\ \mathrm{m.s^{-1}}$ and $h=3\ \mathrm{cm}$.}}
	\label{setup_layout}
\end{figure}

The present investigation was conducted using the experimental setup sketched in figure \ref{setup_layout}(a). It consists of a glass tank \WS{of length 2 m, width 15 cm and height 30 cm}, in which is placed an aluminum vertical wall \WS{of thickness 1 cm, width 14 cm and height 35 cm}, referred to as the piston in the following. An additional aluminum framing is fixed to the back of the vertical wall to prevent it from deforming when it is set into motion and to avoid oscillations when stopped. In addition, a rubber seal is glued to the lateral sides of the piston to maximize tightness. The piston is initially positioned at one end of the flume and is connected to a linear brushless servo-motor (Transtechnik DSM 5.22.11Z8).

At the beginning of a series of experiments, water is poured into the flume up to a height $h$ which defines the initial water depth, as illustrated in figure \ref{setup_layout}(a). The $x$ axis is defined along the streamwise direction of the channel, the $y$ axis follows the spanwise direction and the $z$ axis is oriented vertically and opposed to gravity. The origin is set at the undisturbed water level, at the initial contact between the piston and the liquid. Water is dyed with fluorescein to enhance the contrast. The piston is then translated along the $x$ axis at a controlled velocity $v_p$ and stroke $x_p$. As a result of its translating motion, a certain amount of water is displaced, thereby generating an impulse wave. Two parameters set the dynamics of the piston: the maximum velocity $U$ reached by the translating wall during its horizontal course and its total stroke $L$, \textit{i.e.}, its final position along the channel. It should be emphasised that $U$ and $L$ can be chosen independently with the present experimental setup. The \WS{prescribed} impulsive forcing motion is \WS{the following}: the piston first undergoes a constant acceleration phase where its position is given by 

\begin{equation}
	\label{acceleration_law}
	x_p(t)=\frac{U^2}{2L}t^2 \quad \mathrm{\ for\ } 0 \leqslant t \leqslant L/U,
\end{equation}

\noindent followed by a constant deceleration where

\begin{equation}
	\label{deceleration_law}
	x_p(t)=L \left(- \frac{U^2}{2L^2} t^2 + 2\frac{U}{L}t - 1 \right) \quad \mathrm{\ for\ } L/U \leqslant t \leqslant 2L/U.
\end{equation}

\noindent As a result, for both stages, the \WS{prescribed} evolution of the position of the piston is quadratic in time, whereas the corresponding velocity evolves linearly, as illustrated in figure \ref{setup_layout}(b) \WS{by the solid and dashed lines}, \WS{ respectively}. By doing so, the \WS{prescribed} wave forcing is symmetrical with a constant acceleration $\gamma = U^2/L$ (respectively, deceleration $-U^2/L$) during the first (respectively, second) phase of the motion. The choice of this ``free-fall''-like law of motion is motivated by its geophysical relevance, as highlighted by recent studies on dry granular collapses and subsequent generated impulse waves, where the granular front was found to behave in a similar manner \citep{2021c_sarlin,2022a_sarlin}. \WS{A typical measurement of the effective motion of the piston during an experiment in terms of relative stroke (\textcolor{color_curve_xp}{$\square$}) and velocity (\textcolor{color_curve_vp}{$\circ$}) is shown in figure \ref{setup_layout}(b). In all cases, the recorded motion is observed to be very close to the prescribed one.}

Through the experiments, the initial water depth $h$ was varied in the range $[1 , 23]$ cm, the stroke $L$ of the piston between 2 cm and 30 cm and its maximum velocity $U$ in the range $[0.1 , 1.2 ]\ \mathrm{m.s^{-1}}$. From these parameters, we define two dimensionless numbers that are the relative stroke of the piston, $\Lambda_p=L/h$, varied here between 0.1 and 10, and the Froude number, $\mathrm{Fr}_p=U/\sqrt{gh}$, \WS{based on the maximum velocity $U$ of the piston and} varied in the range $[0.09, 2.2]$. The systematic variation of these parameters leads to a data set of 266 experiments, which substantially extends the phase space covered by previous studies \citep{1966_miller,1972_das}.

A Nikon D3300 camera, operating at $50$ Hz, records the wave generation process from the side of the glass tank. \WS{As a result, the measurements of the free surface elevation correspond to the liquid height in the vicinity of the side wall}. Based on preliminary experiments, the camera is placed so as to fully capture the formation stage of the wave and the first moments of its propagation along the channel, with a spatial resolution varying approximately between $0.2$ mm and $0.7$ mm. A set of custom \textsc{MATLAB} routines, based on a thresholding method, allows us to extract the water free surface elevation, $\eta(x,t)$, from the video recordings. From there, the amplitude \WS{$A_0(t) =\eta(x_p,t)$} at the contact with the moving wall is determined, as well as the corresponding mid-height width $\lambda_0$, \WS{defined at any time $t$ from $\eta(x_p+\lambda_0,t)=A_0(t)/2$}. To check the reproducibility of the generated waves, several representative experiments were repeated five times. By doing so, the wave characteristics $A_0$ and $\lambda_0$ obtained at $t=L/U$ were found to vary by less than 1 \% and 4 \%, respectively. \WS{We did not observe any significant variation in height across the span of the channel, as reported in \citet{Sarlin2022c} where views from a different angle are provided.}


\section{Mapping of the wave regimes}
\label{sec_wave_regimes}

\subsection{Phase diagram of the generated impulse waves}
\label{subsec_phase_diag}

\begin{figure}
	\centerline{\includegraphics[width=\linewidth]{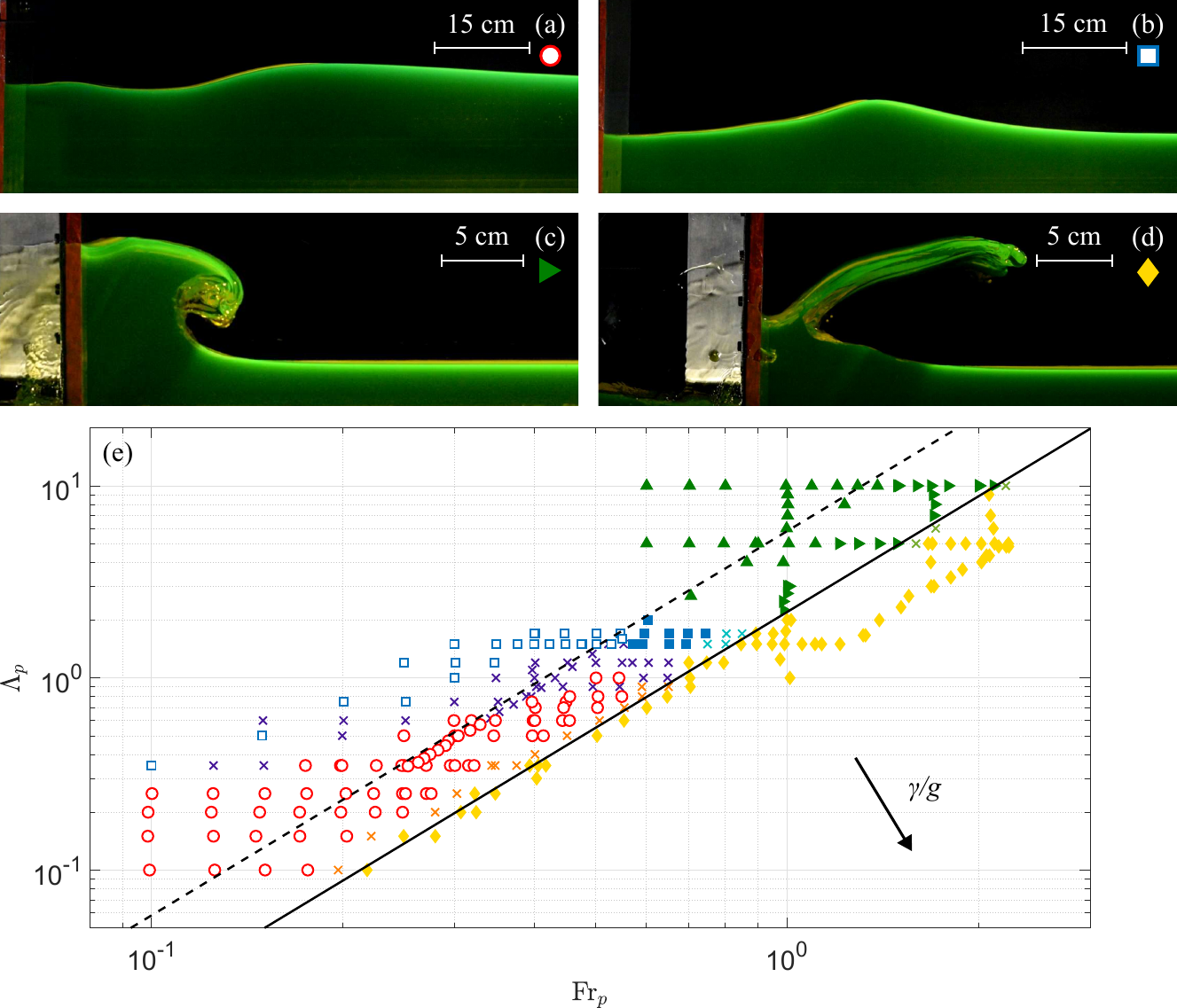}}
	\caption{Representative cases of the different regimes of impulse waves observed: (a) a dispersive wave for $L=7\ \mathrm{cm}$, $U=0.42\ \mathrm{m.s^{-1}}$, and $h=20\ \mathrm{cm}$ \WS{($\mathrm{Fr}_p = 0.3$ and $\Lambda_p = 0.35$) at time $t = 0.72\ \mathrm{s} \simeq 4.3 L/U$ after the beginning of the piston movement}, (b) a solitary-like wave for $L=15\ \mathrm{cm}$, $U=0.47\ \mathrm{m.s^{-1}}$, and $h=10\ \mathrm{cm}$ \WS{($\mathrm{Fr}_p = 0.47$ and $\Lambda_p = 1.5$) at time $t = 0.76\ \mathrm{s} \simeq 2.4 L/U$}, (c) a plunging breaking bore for $L=30\ \mathrm{cm}$, $U=1.09\ \mathrm{m.s^{-1}}$, and $h=3\ \mathrm{cm}$ \WS{($\mathrm{Fr}_p = 2.0$ and $\Lambda_p = 10$) at time $t = 0.36\ \mathrm{s} \simeq 1.3 L/U$}, and (d) a water jet for $L=14.5\ \mathrm{cm}$, $U=1.19\ \mathrm{m.s^{-1}}$, and $h=3\ \mathrm{cm}$ \WS{($\mathrm{Fr}_p = 2.2$ and $\Lambda_p = 4.8$) at time $t = 0.3\ \mathrm{s} \simeq 2.5 U/L$}. For each experiment, the corresponding scale is indicated by a white bar. (e) Diagram of the impulse wave regimes in the $(\mathrm{Fr}_p,\Lambda_p)$ plane: (\raisebox{0.15mm}{\textcolor{color_pwd_1}{$\circ$}}) dispersive waves, (\scalebox{0.9}{\textcolor{color_pwd_2}{$\square$}}) nonbreaking and (\scalebox{0.9}{\textcolor{color_pwd_2}{$\blacksquare$}}) breaking solitary\WS{-like} waves, (\textcolor{color_pwd_4}{$\blacktriangle$}) spilling and (\textcolor{color_pwd_5}{$\blacktriangleright$}) plunging breaking bores, and (\scalebox{1}{\textcolor{color_pwd_6}{$\blacklozenge$}}) water jets, respectively. \WS{Crosses ($\times$)} correspond to experiments at the transition between several wave types, for which it is not straightforward to discriminate between regimes. The solid line (---) marks the transition from waves to water jets when $\Lambda_p = 2.2\, {\mathrm{Fr}_p}^2$ or $U^2/L \simeq 0.45\,g$. The dashed line (\dashedblack) corresponds to the expression $\Lambda_p = 5.8\, {\mathrm{Fr}_p}^2$ (or equivalently $U^2/L \simeq 0.17\,g$), observed in the context of impulse waves engendered by the collapse of a granular column in shallow water \citep{2022a_sarlin}.}
	\label{wave_regimes_diagram}
\end{figure}

Various kinds of waves are observed through the experiments, as illustrated in figure \ref{wave_regimes_diagram}, which suggests a great behavioural richness. These observations echo the different regimes previously reported in the context of wave generation by a rigid wall \citep{1966_miller,1970_noda,1972_das} or by the entrance of a Newtonian fluid \citep{2022_kriaa} or a granular material \citep{2004_fritz,2011_heller,2021b_sarlin} into water. When both the stroke and velocity of the piston are small enough (\textit{i.e.}, for $L<h$ and $U < \sqrt{gh}$), dispersive waves are generated [see figure \ref{wave_regimes_diagram}(a)], which are akin to damped oscillations featuring a decrease in amplitude over time and a frequency dispersion during their propagation. The behaviour of these waves, and especially the evolution of their free-surface elevation, is reminiscent of that of the Cauchy--Poisson waves discussed, for instance, by \citet{1999_whitham}. Dispersion gets balanced by non-linearity when the stroke of the piston is increased ($L \sim h$ and $U < \sqrt{gh}$), leading to the formation of solitary\WS{-like} waves as illustrated in figure \ref{wave_regimes_diagram}(b). In such cases, the sole propagating peak is not necessarily stable as, in some experiments, the wave eventually breaks. The presence of \WS{such solitary-like waves} indicates that the law of motion of the piston approaches here the particular situations investigated in the studies of \citet{1980_goring}, \citet{1990_synolakis} and \citet{2002_guizien}, where these authors solved an inverse problem to accurately generate experimentally solitary waves. When $L > h$, bore waves are engendered and present a characteristic non-linear steepening of the wavefront that systematically leads to wave breaking as illustrated in figure \ref{wave_regimes_diagram}(c). \WS{Wave breaking is spilling when $U$ is close to $\sqrt{gh}$, so that no noticeable air entrapment is visible. By increasing $U$ while all other parameters are kept fixed, breaking} occurs increasingly closer to the piston, as non-linear effects become dominant. This eventually leads to the formation of plunging breakers for a sufficiently large velocity of the piston ($U > \sqrt{gh}$), for which a significant air entrapment by the breaking wave is observed \WS{[see figure \ref{wave_regimes_diagram}(c)]}. Finally, a last situation is encountered experimentally when the typical acceleration of the piston is large enough: in this case, a peculiar thin jet of water gets propelled downstream from the forming wave, as illustrated in figure \ref{wave_regimes_diagram}(d). \WS{This regime seems to mark the transition from classical waves to splashes and,} to the best of our knowledge, this is the first report of such a fluid structure in the context of wave generation by a translating wall. \WS{These water jets are, however,} reminiscent of the ``hydrodynamic impact craters'' reported by \citet{2003b_fritz,2003c_fritz} when these authors studied the formation of impulse waves caused by the impact of a thin granular slide \WS{propelled pneumatically}. In our case, once the jet detaches from the vicinity of the piston, it seems to experience a ballistic motion over its course before impacting the free surface of the main fluid body, thereby dissipating a lot of energy in the process. The water jet is reproducible, \textit{i.e.}, repeating the same experiment twice results in the same observed fluid motion \WS{and free surface deformation}. \WS{The video recordings of the four cases presented in figures \ref{wave_regimes_diagram}(a)-(d) are available in supplementary material, alongside a typical experiment of a spilling breaking bore.}
It should be emphasised that all generated waves can be considered as gravity waves, as they exhibit a longitudinal dimension significantly larger than the capillary wavelength of water, $\lambda_c=2\pi (\sigma/(\rho g))^{1/2} \simeq 1.7$ cm (where $\sigma=72\ \rm{mN.m^{-1}}$ and $\rho=997\ \rm{kg.m^{-3}}$ are the water surface tension and density evaluated at a temperature of $25$ \textcelsius, respectively). The only exception is the water jet regime for which, in some cases, the developing fluid filament is very thin, as illustrated in figure \ref{wave_regimes_diagram}(d). As a result, it is possible that some parts of it are affected by surface tension \WS{after the generation process}. However, we will not focus on \WS{this situation} in the following discussion.

Following these observations, the parameters $h$, $L$ and $U$ have been systematically varied, in order to constitute a phase map in the (\Fp,\ \Lp) plane that is represented in figure \ref{wave_regimes_diagram}(e). \WS{For each experiment}, the corresponding wave regime \WS{has been determined by visual inspection of the video recordings and is} reported with distinct symbols for (\raisebox{0.15mm}{\textcolor{color_pwd_1}{$\circ$}}) dispersive waves, (\scalebox{0.9}{\textcolor{color_pwd_2}{$\square$}}) nonbreaking and (\scalebox{0.9}{\textcolor{color_pwd_2}{$\blacksquare$}}) breaking solitary\WS{-like} waves, (\textcolor{color_pwd_4}{$\blacktriangle$}) spilling and (\textcolor{color_pwd_5}{$\blacktriangleright$}) plunging breaking bores, and (\scalebox{1}{\textcolor{color_pwd_6}{$\blacklozenge$}}) water jets. In the diagram, \WS{crosses} ($\times$) correspond to experiments at the transition between regimes, for which it is challenging to objectively discriminate. \WS{No data are available in the bottom right-hand corner of figure \ref{wave_regimes_diagram}(e) as this is the realm of large relative accelerations $\gamma/g \equiv U^2/(gL)$, whereas the motorized piston used in the present study could only reach $\gamma/g \simeq 1$. Furthermore, no experiments have been reported in the top left-hand corner of figure \ref{wave_regimes_diagram}(e) corresponding to low relative acceleration, typically $\gamma/g \lesssim 0.04$ as, in this case, the length of our channel is too small to be able to clearly distinguish the wave regime.} As highlighted by the solid line \WS{of slope 2 in the log-log plot} of figure \ref{wave_regimes_diagram}(e), \WS{the transition from waves to} water jets \WS{(or splashes)} occurs when $\Lambda_p \lesssim 2.2\,{\mathrm{Fr}_p}^2$, or equivalently

\begin{equation}
\label{jet_criteria}
\frac{\gamma}{g} = \frac{U^2}{gL} \gtrsim 0.45.
\end{equation}



\noindent Interestingly, this criterion does not constitute a threshold based on $\Lambda_p$ or \Fp\ alone, but it combines the two parameters. From there, it can be inferred that, for a given value of \WS{the relative forcing length} $\Lambda_p$, one can produce such a water jet by starting from either of the three previously described regions (\textit{i.e.}, of dispersive, solitary\WS{-like} or bore waves) and then increasing the Froude number up to the point where relation \eqref{jet_criteria} is satisfied. \WS{A dashed line, corresponding to the relation} $\gamma/g \simeq 0.17$ ($\Lambda_p = 5.8\, {\mathrm{Fr}_p}^2$), is also reported in figure \ref{wave_regimes_diagram}(e). This particular value for the relative acceleration \WS{is observed in the context of} impulse waves \WS{triggered by gravity-driven granular collapses in shallow water, where the granular front acts like a moving piston \citep{2022a_sarlin}}. Such a relation between $\gamma$ and $g$ reveals that the existing coupling between the position of the granular front and its velocity inherently selects the possible wave regimes observed during a granular collapse into water (either dispersive, solitary, or bore waves) \WS{and explains why no water jets could be observed by \citet{2021b_robbe-saule} and \citet{2021b_sarlin}}. \WS{There is a qualitative agreement between the results reported in the present study and those obtained by \citet{1972_das}, who used a similar experimental configuration. However, a direct quantitative comparison is not straightforward because the Froude number considered by these authors is based on the average velocity of the piston rather than on the maximum velocity used here while, at the same time, the precise law of motion of the piston is not provided by them. The main difference lies in the fact that \cite{1972_das} did not observe the water jet regime, probably because their setup did not allow them to achieve a high enough relative acceleration.}


\subsection{The observed waves regimes in light of the Korteweg -- de Vries equation}
\label{subsec_kdv}

The richness of the physics at play explains the long-standing research interest on how to relate each kind of wave to an existing wave theory \citep{2004_fritz,2011_heller}. However, a simplified analysis might already be helpful in the aim of understanding the origins of these different regimes of impulse waves. Indeed, in the case of weakly non-linear shallow water waves, one can consider, for instance, the classical Korteweg -- de Vries equation to describe the evolution of the free surface elevation, $\eta(x,t)$ \citep{1895_korteweg}. In its standardized dimensionless form, this equation reads





\begin{equation}
\label{eqkdv1}
\eta^*_{t^*} + \eta^* \eta^*_{x^*} + \eta^*_{x^*x^*x^*} =0,
\end{equation}

\noindent with $t^* \equiv t \sqrt{g/h} / 6$, $x^* \equiv x/h$, and $\eta^* \equiv 9 \eta/h + 6$. Equation \eqref{eqkdv1} involves a competition between a non-linear term, $\eta^* \eta^*_{x^*}$, that tends to steepen the wavefront, and a dispersive one, $\eta^*_{x^*x^*x^*}$, which promotes the appearance of an oscillatory behaviour \citep{1999_whitham,2004_dauxois}. In a slightly different approach, \WS{we} consider \WS{here} an alternate transformation based on the variables $\tilde t \equiv t \sqrt{g/h} / 6$, $\tilde x \equiv x/L$, and $\tilde \eta \equiv 9 \eta/h + 6$, which leads to




\begin{equation}
\label{eqkdv2}
\tilde \eta_{\tilde t} + {\Lambda_p}^{-1} \, \tilde \eta \tilde \eta_{\tilde x} + {\Lambda_p}^{-3} \, \tilde \eta_{\tilde x \tilde x \tilde x}=0.
\end{equation}

\noindent This writing, which differs from the previous one by using $L$ instead of $h$ for making $x$ dimensionless, reveals the importance of the relative stroke of the piston, $\Lambda_p=L/h$, in the selection of the wave regime. Indeed, equation \eqref{eqkdv2} suggests that increasing $\Lambda_p$ tends to favor non-linearity, which is in agreement with the presence of bore waves (\textcolor{color_pwd_5}{$\blacktriangleright$}), (\scalebox{1}{\textcolor{color_pwd_6}{$\blacklozenge$}}) for large values of this dimensionless number [figure \ref{wave_regimes_diagram}(e)]. Contrariwise, decreasing $\Lambda_p$ enforces dispersive effects and, as a result, the development of oscillatory waves (\raisebox{0.15mm}{\textcolor{color_pwd_1}{$\circ$}}), which also agrees with the present experimental observations. This analysis also suggests that the Froude number has \WS{a less} significant influence on the selection of the wave regime, as illustrated in figure \ref{wave_regimes_diagram}(e). \WS{Indeed, it can be observed at a crude first order that, by increasing $\mathrm{Fr}_p$ while keeping a constant value for $\Lambda_p$, there is almost no change from a wave regime to another above the transition from waves to water jets delimited by equation \eqref{jet_criteria}}. Nevertheless, \WS{the Froude number} strongly determines whether the generated waves are stable or not: for instance, as can be seen in figure \ref{wave_regimes_diagram}(e), \WS{empty (\textit{i.e.}, non-breaking) and filled (\textit{i.e.}, breaking) symbols for the solitary-like and bore wave regimes are separated based on $\mathrm{Fr}_p$ in such a way that} breaking systematically occurs when $\mathrm{Fr}_p \gtrsim 0.6$, regardless of the value of $\Lambda_p$. \WS{However, all these observations do not apply to the water jet regime, which occurs for relative accelerations larger than the critical value given by equation \eqref{jet_criteria}, as discussed previously.}

The presence of these various wave regimes raises questions about their generation process: is there a universal manner of describing it, or are different approaches necessary to this end? What governs the typical extent of the disturbance produced by the translational motion of the piston when it injects energy into the fluid? To address these points, a more thorough analysis of the wave hydrodynamics during the generation phase is provided thereafter. 


\section{The birth of an impulse surface wave}
\label{sec_generation}


\subsection{Time evolution of the induced water bump}
\label{subsec_bump}

\begin{figure}
	\centerline{\includegraphics[width=0.98\linewidth]{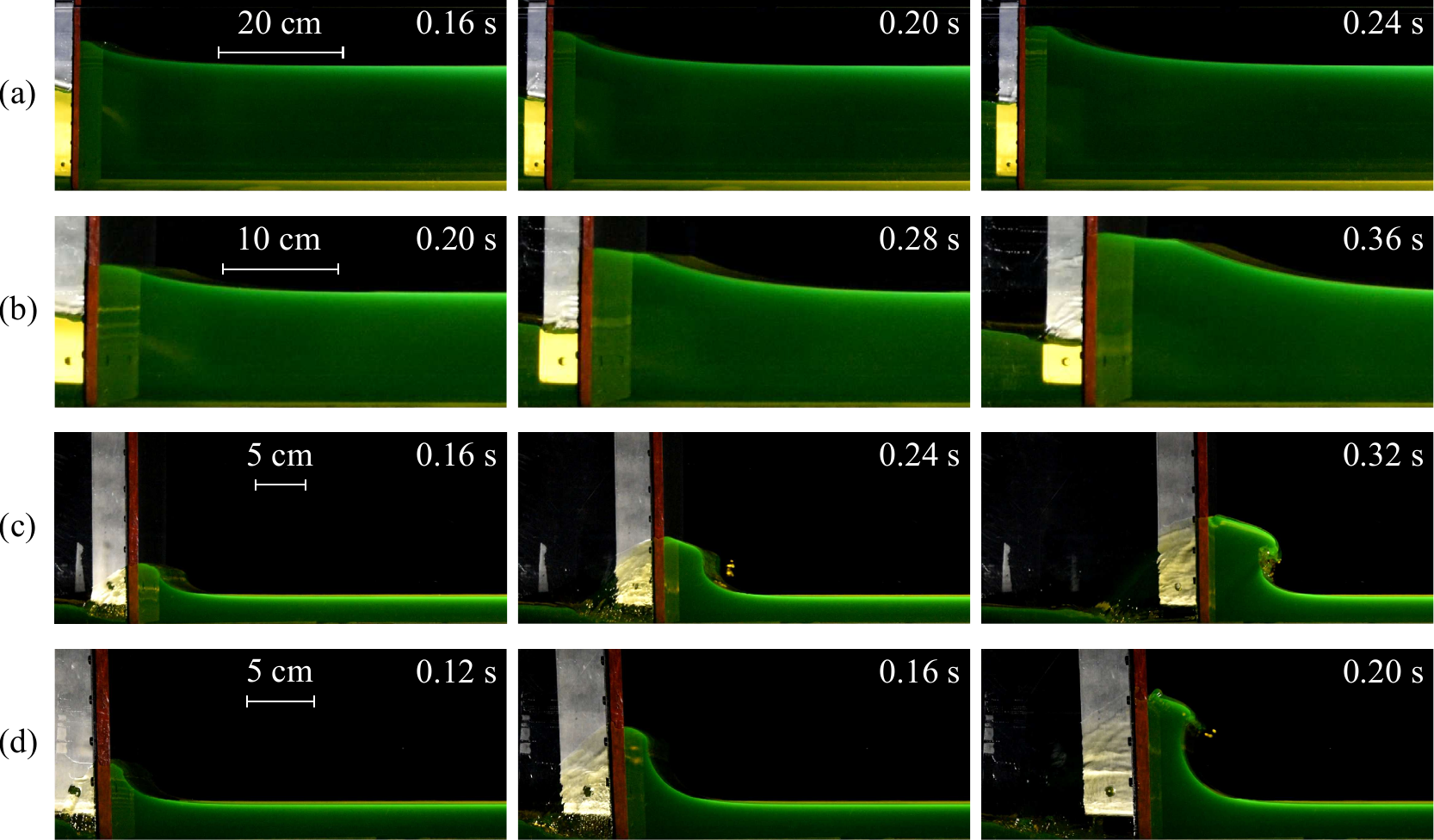}}
	\caption{Image sequences of the generation phase, for the impulse waves presented in figure \ref{wave_regimes_diagram}: (a) $L=7\ \mathrm{cm}$, $U=0.42\ \mathrm{m.s^{-1}}$, and $h=20\ \mathrm{cm}$ ($\mathrm{Fr}_p = 0.3$ and $\Lambda_p = 0.35$); (b) $L=15\ \mathrm{cm}$, $U=0.47\ \mathrm{m.s^{-1}}$, and $h=10\ \mathrm{cm}$ ($\mathrm{Fr}_p = 0.47$ and $\Lambda_p = 1.5$); (c) $L=30\ \mathrm{cm}$, $U=1.09\ \mathrm{m.s^{-1}}$, and $h=3\ \mathrm{cm}$ ($\mathrm{Fr}_p =2.0$ and $\Lambda_p = 10$); (d) $L=14.5\ \mathrm{cm}$, $U=1.19\ \mathrm{m.s^{-1}}$, and $h=3\ \mathrm{cm}$ ($\mathrm{Fr}_p = 2.2$ and $\Lambda_p = 4.8$). The last picture of each line is taken at the end of the generation phase, \textit{i.e.}, when the wave amplitude at the contact with the piston reaches its maximum value $A_{m0}$. For each experiment, the white bar traced on the first image gives the scale.}
	\label{bump_rise}
\end{figure}

When the piston starts its translational motion along the channel, a water bump forms in the vicinity of the advancing wall, leaving the rest of the fluid undisturbed. This initial perturbation then grows in volume as long as the vertical wall is accelerating, as illustrated in figure \ref{bump_rise} for four representative experiments. At the end of this generation phase, the bump reaches its maximum elevation $A_{m0}$ at the contact with the piston. After this moment, \textit{i.e.}, during the deceleration of the wall, the local water disturbance detaches from the piston and relaxes, thereby evolving to a wave that propagates away from the source region and belongs to one of the regimes previously described in figure \ref{wave_regimes_diagram}. The shape of the growing water bump varies between the investigated configurations, as highlighted in figure \ref{bump_rise}. Broadly speaking, the perturbation has a small amplitude and a large width for a small velocity of the piston [figures \ref{bump_rise}(a)-(b)], whereas it has the shape of a slim water column for large values of $U$ [figures \ref{bump_rise}(c)-(d)].

\begin{figure}
	\centerline{\includegraphics[width=\linewidth]{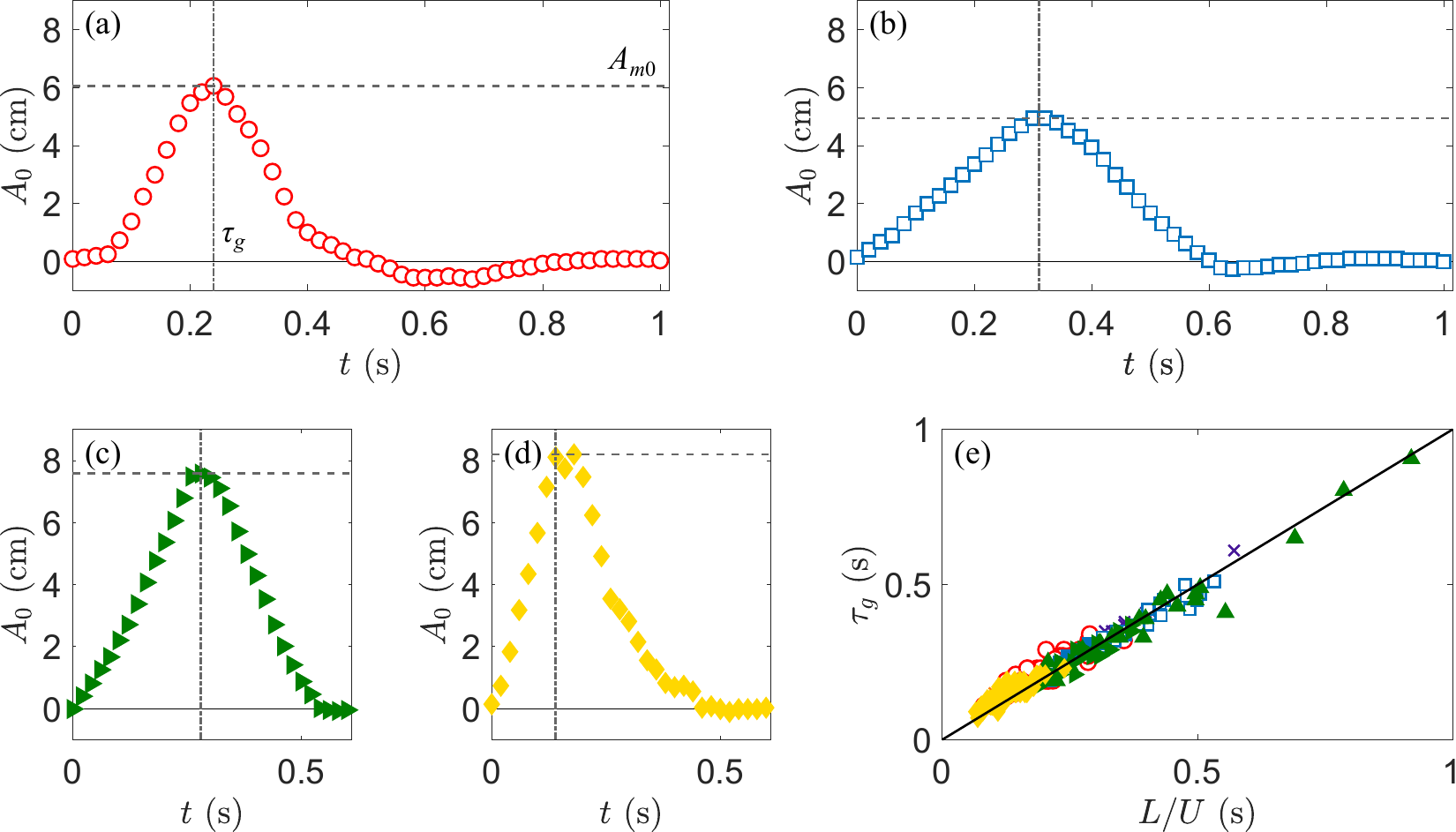}}
	\caption{(a)-(d) Temporal evolution of the free surface elevation $A_0(t)=\eta(x=x_p(t),t)$ at the contact with the piston, for the four experiments of figure \ref{bump_rise}. (a) $L=7\ \mathrm{cm}$, $U=0.42\ \mathrm{m.s^{-1}}$, and $h=20\ \mathrm{cm}$, (b) $L=15\ \mathrm{cm}$, $U=0.47\ \mathrm{m.s^{-1}}$, and $h=10\ \mathrm{cm}$, (c) $L=30\ \mathrm{cm}$, $U=1.09\ \mathrm{m.s^{-1}}$, and $h=3\ \mathrm{cm}$, and (d) $L=14.5\ \mathrm{cm}$, $U=1.193\ \mathrm{m.s^{-1}}$, and $h=3\ \mathrm{cm}$. For each case, the vertical dash-dotted line indicates the generation time, $\tau_g$, whereas the horizontal dashed line corresponds to the maximum wave amplitude $A_{m0}$ at the junction between the fluid and the advancing wall. (e) Generation time $\tau_g$ as a function of the duration $L/U$ of the acceleration phase of the piston. The solid line corresponds to $\tau_g=L/U$. In (a)-(e), the symbols and colours used are the same as in figure \ref{wave_regimes_diagram}(e).}
	\label{A0_t}
\end{figure}

The temporal evolution of the wave amplitude $A_0(t)=\eta(x=x_p,t)$ at the contact with the piston is illustrated in figure \ref{A0_t}(a)-(d) for the four experiments reported in figure \ref{bump_rise}. For every configuration, $A_0$ increases with time until it reaches a maximum value $A_{m0}$ at time $t=\tau_g$. $A_0$ then decreases when the moving wall decelerates and the wave begins to propagate along the channel. At later times, when the wave has left the vicinity of the piston, $A_0$ tends to zero as the water locally comes back to rest: this is illustrated, for instance, in figure \ref{A0_t}(b) when $t \gtrsim 0.6$ s. Noticeably, all curves display the same overall behaviour, regardless of the experimental parameters. The bell-shaped trends for $A_0$ are not perfectly symmetrical about the vertical line $t=\tau_g$, which suggests that the two phases of generation and propagation are driven by different physical mechanisms. Indeed, during the first stage, the translating wall injects momentum into the fluid, but once the wave travels into the channel it does not receive energy anymore, so that its evolution is then described by a competition between dissipation, non-linearity and dispersion. The generation time $\tau_g$, defined as the duration of the growth phase of $A_0$, is systematically extracted for all experiments and compared in figure \ref{A0_t}(e) to the duration of the acceleration phase of the piston, $L/U$. Overall, it can be observed that the two times coincide for the whole dataset. The scattering of the data is mainly due to the experimental uncertainty in determining $\tau_g$, which is of order $0.03$ s, however no systematic deviation can be observed. This result indicates that the wave generation process is closely tied to the acceleration phase of the translating piston, so that $L/U$ is the relevant timescale for the generation stage.

\begin{figure}
	\centerline{\includegraphics[width=\linewidth]{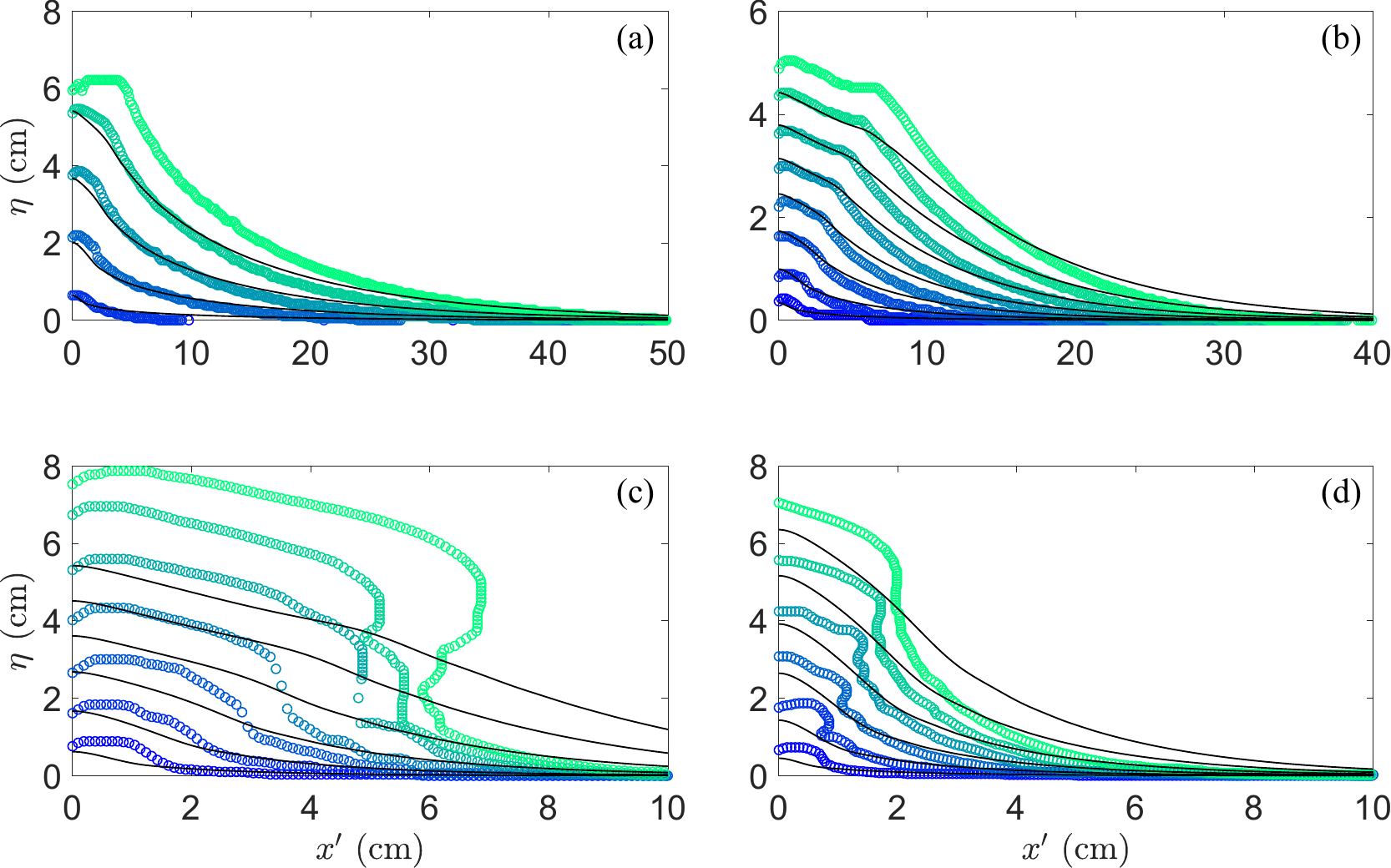}}
	\caption{Transient free surface elevation $\eta$ during the generation phase as a function of the distance $x^\prime$ from the piston for the four examples of figure \ref{bump_rise}\WS{, corresponding to relative accelerations $\gamma/g$ of (a) 0.26, (b) 0.15, (c) 0.4 and (d) 1, respectively}. Experiments are represented by the coloured markers, with a timelapse between each consecutive curve of (a)-(c) $0.04$ s and (d) $0.02$ s. The solid lines are the corresponding predictions from the theory of \citet{1990_joo}, given by equation \eqref{joo_solution_dim} with $\sigma=72\ \rm{mN.m^{-1}}$, $\rho=997\ \rm{kg.m^{-3}}$ and assuming a contact angle of $90^\circ$. Here, $x^\prime = x-x_p(t)$ for the experimental curves, while $x^\prime = x$ for the theoretical profiles.}
	\label{transient_bump_growth}
\end{figure}

To further elaborate on the description of the water bump formation, the free surface elevation $\eta(x,t)$ is presented in figure \ref{transient_bump_growth}(a)-(d) as a function of the distance $x'=x-x_p(t)$ from the vertical wall, for the representative experiments presented in figure \ref{bump_rise}. On each plot, several successive profiles are reported (one for each marker colour), with a time increment between two consecutive curves of $0.04$ s in (a)-(c) and $0.02$ s in (d). In all cases, one may note that both the height and the horizontal extent of the water perturbation increase with time. In figure \ref{transient_bump_growth}(a), $\Lambda_p<1$ and $\mathrm{Fr}_p<1$, which results in a short bump with a small vertical extent compared to the large horizontal length affected by the disturbance. At all times, the elevation of the free surface in this situation essentially exhibits a decay with the distance from the piston, except at the very vicinity of the advancing wall. This decrease of the free surface elevation with $x^\prime$ is reminiscent of the logarithmic decay at small times identified by previous theoretical studies \citep{1983_chwang,1984_lin}. When $\Lambda_p \sim 1$ while $\mathrm{Fr}_p<1$, similar characteristics are observed, as illustrated in figure \ref{transient_bump_growth}(b), but the decrease of $\eta$ with $x'$ is now divided into two regions: a relatively small slope in the vicinity of the piston, which is less pronounced than the one occurring further downstream [highlighted, for instance, by the evolution of $\eta$ after $x' \simeq 7$ cm for the upper green curve in figure \ref{transient_bump_growth}(b)]. This phenomenon gets accentuated when $\Lambda_p \gg 1$ and $\mathrm{Fr}_p \sim 1$, as illustrated in figure \ref{transient_bump_growth}(c): the region of gentle slope then becomes wider. In the upper green curve of figure \ref{transient_bump_growth}(c), corresponding to the last time belonging to the generation phase, $\eta$ decreases by only about 2 cm over a distance of $6$ cm in the inner region \WS{located upstream of the shock}. A non-linear steepening occurs before the end of the generation process, leading to the formation of a hydrodynamic shock, that is illustrated by the onset of breaking visible in figure \ref{bump_rise}(c) for $t=0.32$ s or by the upper green curve in figure \ref{transient_bump_growth}(c) which features a straight front around $x' \sim 6$ cm. Finally, by reducing the piston stroke $L$ while keeping the same initial water depth $h$ and velocity $U$ of the piston as in figure \ref{transient_bump_growth}(c), one obtains a tall water bump, as illustrated in figure \ref{transient_bump_growth}(d), which will evolve to a water jet after the generation phase [figure \ref{wave_regimes_diagram}(d)]. In that case, the gentle slope region has a smaller extent than the outer one, and the aspect ratio of the growing perturbation becomes significant.

\subsection{Volume and aspect ratio of the \WS{water} bumps}
\label{subsec_generation_time_and_shape}
 
At the end of the generation phase \WS{(at $\tau_g \simeq L/U$)}, the two characteristic lengthscales associated with the induced water hump can be taken as the maximum amplitude $A_{m0}$ at the contact with the piston and the mid-height width $\lambda_{m0}$, which is defined as the width of the perturbation at $z=A_{m0}/2$ \WS{so that $\eta(x_p+\lambda_{m0},\tau_g)=A_{m0}/2$}. These two quantities define a typical bump volume per unit width, $A_{m0} \lambda_{m0}$. By mass conservation, it is straightforward to establish that $A_{m0} \lambda_{m0} = \vartheta Lh$, where $\vartheta$ is a numerical prefactor that depends on the shape of the forming wave \WS{and on the leaks between the flume walls and the piston that lower the displaced volume of water}. By making this relation dimensionless with the use of the lenghtscale $h$, one obtains

\begin{equation}
\label{lambda_h_1}
\frac{A_{m0} \lambda_{m0}}{h^2} = \vartheta \Lambda_p.
\end{equation}

\begin{figure}
	\centerline{\includegraphics[width=\linewidth]{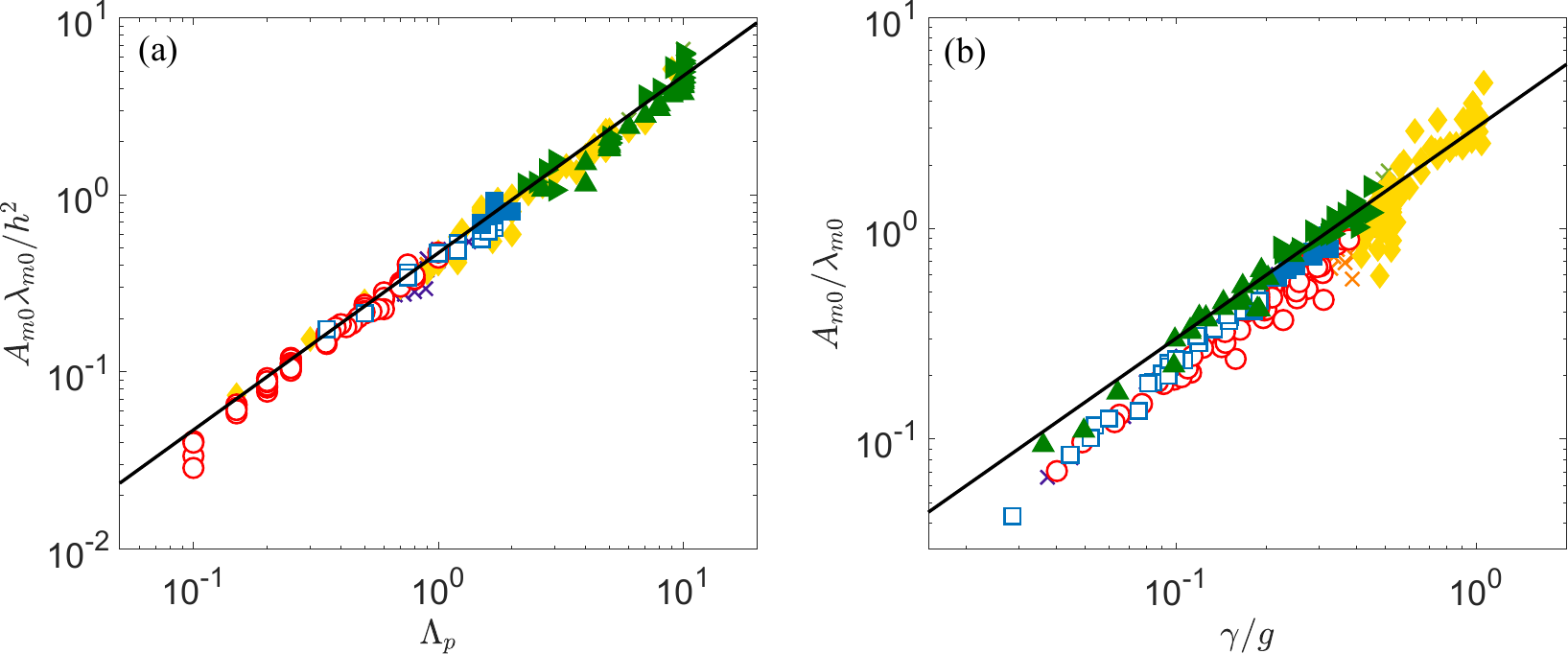}}
	\caption{(a) Dimensionless water bump volume $A_{m0}\lambda_{m0}/h^2$ per unit width at the end of the generation phase as a function of the relative piston stroke $\Lambda_p$. (b) Evolution of the bump aspect ratio $A_{m0}/\lambda_{m0}$ with the relative acceleration $\gamma/g$ of the piston. In both plots, all experiments are represented and the solid line indicates (a) $A_{m0}\lambda_{m0}/h^2=0.47\,\Lambda_p$ and (b) $A_{m0}/\lambda_{m0}=3\,\gamma/g$. The symbols and colours used are the same as in figure \ref{wave_regimes_diagram}(e).
	}
	\label{wave_volume_and_aspect_ratio}
\end{figure}

\noindent In figure \ref{wave_volume_and_aspect_ratio}(a), the dimensionless bump volume per unit width, $A_{m0} \lambda_{m0}/h^2$, is reported as a function of the relative stroke of the piston, $\Lambda_p$, for all experiments. All data collapse on the master curve $A_{m0} \lambda_{m0}/h^2 = 0.47 \Lambda_p$ revealing that, at leading order, the typical volume per unit width $A_{m0} \lambda_{m0}$ of the nascent wave is proportional to $Lh$ with no significant influence of the \WS{details of the} bump shape through the numerical prefactor $\vartheta$. In other terms, when $L$ and $h$ are set to a fixed value, the higher the maximum amplitude $A_{m0}$, the smaller the characteristic length $\lambda_{m0}$ and vice versa, regardless of the ultimate wave regime obtained after the generation process. Furthermore, this implies that there is solely one relevant lengthscale to describe the generated water bump, which will be taken as $A_{m0}$ in the following.

In addition, the wave aspect ratio $A_{m0}/\lambda_{m0}$ is compared in figure \ref{wave_volume_and_aspect_ratio}(b) with the relative acceleration $\gamma/g=U^2/(gL)$ of the piston. A monotonic increase with $\gamma/g$ is observed: the larger the acceleration of the piston, the slender the resulting water bump. At a crude first order, one can observe the wave aspect ratio to be approximately linear in $\gamma/g$, here again independently of the wave regime obtained at long time. An important remark can be made from these scalings for the displaced volume of water and the aspect ratio of the wave: one can anticipate that $\lambda_{m0}/h$ should be linearly related to the ratio $\Lambda_p/\mathrm{Fr}_p$, whereas the relative amplitude $A_{m0}/h$ should scale with the Froude number $\mathrm{Fr}_p$, at first order. \WS{Nevertheless, a} power law fit on the data displayed in figure \ref{wave_volume_and_aspect_ratio}(b) gives an exponent of 1.23, which \WS{slightly} departs from \WS{such} a linear evolution. Furthermore, if the collapse of the measurement points is quite convincing in figure \ref{wave_volume_and_aspect_ratio}(a), a larger scattering can be noticed in figure \ref{wave_volume_and_aspect_ratio}(b). 

\section{Modelling the wave generation process}
\label{sec_model}

In this section, a particular attention will be devoted to the description of the wave amplitude $A_0(t)$ at the contact with the translating vertical wall, and especially to its maximum value $A_{m0}$ reached at the end of the generation stage (\textit{i.e.}, at the time $L/U$). Given the variety of the observed wave behaviours, it already appears that finding a unique description might be a challenging path. Instead of doing so, we will consider successively the two limiting scenarios of a small relative acceleration $\gamma/g=U^2/(gL)$ of the piston and of a high Froude number $\mathrm{Fr}_p$.

\subsection{Small relative acceleration}
\label{Subsec_model.1}

Describing the impulse wave generation process is a long-standing and challenging problem in the fluid dynamics community, which explains the numerous theoretical studies devoted to the expression of the free surface elevation for various configurations \citep{1984_lin,1984_chwang,1987_roberts,1990_joo,1994_king,2007_needham,2015_uddin}. Among these, \citet{1990_joo} addressed the problem of waves generated by a piston translating horizontally with a uniform acceleration, using a potential flow assumption and taking into account capillary effects due to surface tension and wettability. Using an asymptotic analysis based on a small relative acceleration $\gamma/g$ of the piston, they obtained the following leading-order solution for the free surface elevation

\begin{equation}
\label{joo_solution_dim}
\eta(x,t) = \frac{2h \gamma}{\pi g} \int_0^{+ \infty} \frac{1 - \cos \left( \beta(k,T) t \sqrt{g/h} \right)}{k^2 \left( 1+Tk^2 \right)} \cos \left( k x / h \right) \, \odif{k},
\end{equation}


\noindent where the bound variable $k$ corresponds to a dimensionless wavenumber, $T=\sigma/ \left( \rho g h^2 \right)$ with $\sigma$ the surface tension of the fluid, and $\beta(k,T)=\sqrt{k \left( 1+Tk^2 \right) \tanh k}$.


\noindent Even if expression \eqref{joo_solution_dim} does not constitute a closed-form solution, it can be evaluated numerically. As there is no small-time assumption in the approach followed by \citet{1990_joo}, this model is expected to be more relevant than those of \citet{1984_lin,1984_chwang,1994_king,2007_needham} and \citet{2015_uddin} for a comparison with the present experimental results.




Equation \eqref{joo_solution_dim} is evaluated for the four initial conditions of figure \ref{bump_rise}, up to the time $t=L/U$, and is reported in black solid lines in figure \ref{transient_bump_growth}(a)-(d). In doing so, the surface tension of water is set to $\sigma=72\ \rm{mN.m^{-1}}$ and its density $\rho$ to $997\ \rm{kg.m^{-3}}$ (\textit{i.e.}, their values at a temperature of $25$ \textcelsius), while the contact angle with the wavemaker was assumed to be $90^\circ$. The analytical profiles from equation \eqref{joo_solution_dim} are shown in figure \ref{transient_bump_growth} with a timelapse of (a)-(c) $0.04$ s, and (d) $0.02$ s between each curve. It should be specified that while $x^\prime=x-x_p(t)$ for the experimental curves, $x^\prime = x$ for the theoretical curves. This is due to the fact that \citet{1990_joo} considered the case of a small displacement of the wavemaker, which resulted in neglecting $x_p(t)$ in their analysis while retaining only the influence of the velocity of the piston. An overall good agreement is observed for the first two cases reported in figures \ref{transient_bump_growth}(a) \WS{(for which $\gamma/g \simeq 0.26$)} and \ref{transient_bump_growth}(b) \WS{($\gamma/g \simeq 0.15$)}, corresponding to small relative accelerations of the piston that eventually lead to the formation of (a) a dispersive and (b) a solitary\WS{-like} wave, respectively. This is especially true in the vicinity of the wavemaker, where the analytical predictions are very close to the measured free surface elevations. This is expected as, for both cases, the relative acceleration is $\gamma \ll g$. The observed agreement reveals that, in this case, the theory developed by \citet{1990_joo} gives a quite accurate description of the generated wave, both qualitatively and quantitatively. It should be emphasised that the experimental and theoretical curves are not perfectly superimposed because of the initial time shift present in the video recordings, \WS{that never start exactly at the beginning of the motion of the piston}. Furthermore, one may notice in figure \ref{transient_bump_growth} that the generation stage can last a little longer than $L/U$ for some cases, for there are more experimental curves than theoretical ones. This slight difference could be attributed to second-order entrainment effects. Away from the wavemaker, however, the theoretical predictions deviate from the experimental curves, a feature that is clearly visible in figure \ref{transient_bump_growth}(b), for instance around $x^\prime \simeq 10$ cm for the upper curve. This discrepancy can possibly be attributed to the assumption of a small wavemaker displacement made by \citet{1990_joo} that is not satisfied here. \WS{Finally, as expected, when the relative acceleration becomes significant in the experiments, as illustrated in figures \ref{transient_bump_growth}(c) (where $\gamma/g \simeq 0.4$) and \ref{transient_bump_growth}(d) (for which $\gamma/g \simeq 1$), the leading-order theory of \citet{1990_joo} fails in predicting efficiently the wave generation phase. Indeed, in such cases, non-linear effects become significant, as highlighted for instance by the steepening of the wave front observed during the wave generation for the example in figure \ref{transient_bump_growth}(c)}. As a result, the higher-order terms in the analysis of \citet{1990_joo} can no longer be neglected, hence requiring a dedicated analysis. 


\begin{figure}
	\centerline{\includegraphics[width=\linewidth]{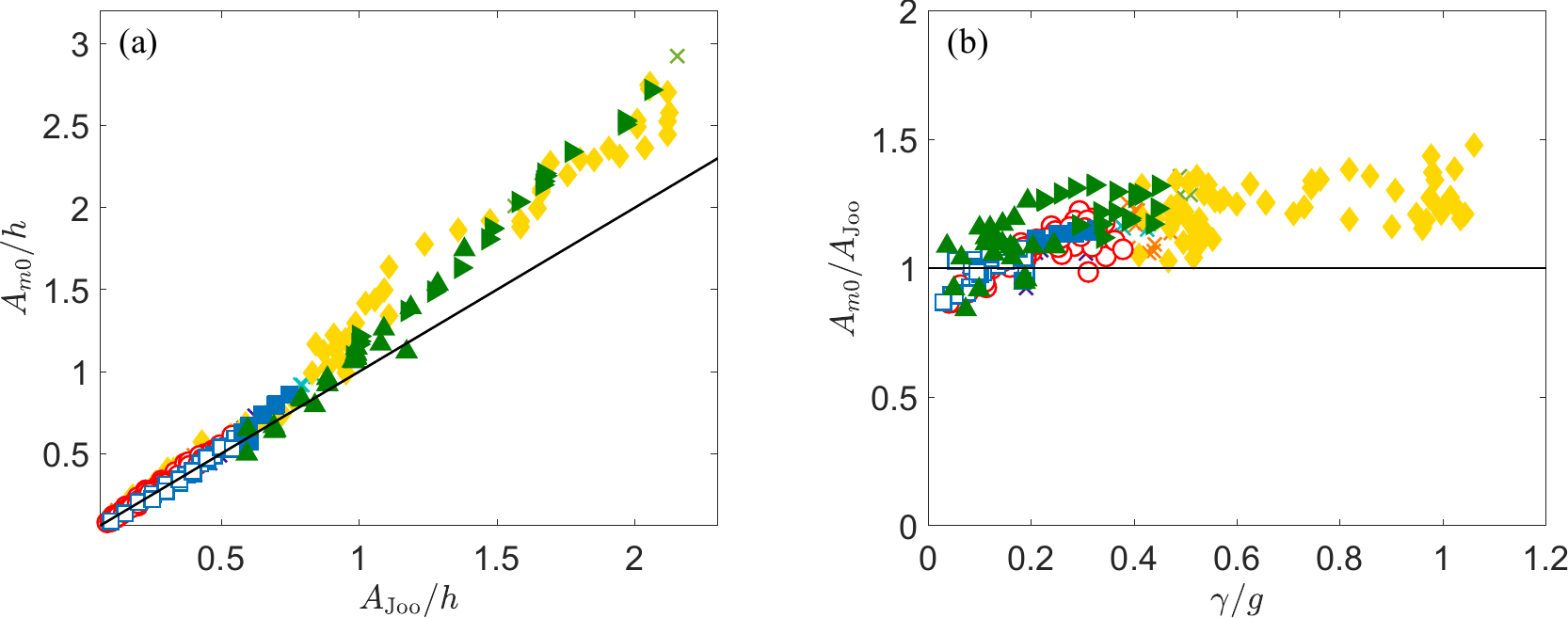}}
	\caption{(a) Comparison between the relative maximum wave amplitude $A_{m0}/h$ measured experimentally and the prediction $A_{\rm{Joo}}/h$ from the theory of \citet{1990_joo}, obtained using equation \eqref{joo_solution_dim}. The solid line is the linear trend of slope one. (b) Evolution of $A_{m0}/A_{\rm{Joo}}$ as a function of the relative acceleration $\gamma/g=U^2/(gL)$. The horizontal solid line is the plateau value $A_{m0}/A_{\rm{Joo}}=1$.
	}
	\label{wave_amplitude_end_generation_joo}
\end{figure}

So far, the region of validity of the theory of \citet{1990_joo} remains to be clarified. In order to discuss this aspect in a more quantitative manner, the comparison between the modelling derived by these authors and the present experimental measurements can be further completed by considering the maximum wave amplitude $A_{m0}$ at the contact with the piston, which is reached at the end of the generation phase. \WS{For each experiment, the corresponding prediction $A_{\rm{Joo}}$ from \citet{1990_joo} is determined by evaluating equation \eqref{joo_solution_dim} for the same initial parameters and at the time $L/U$.} The measured and theoretical values, normalized by the initial liquid depth $h$, are compared in figure \ref{wave_amplitude_end_generation_joo}(a). A good agreement is observed between the two quantities for a large number of water bumps that will ultimately lead to ({\textcolor{color_pwd_1}{$\circ$}}) dispersive or (\scalebox{0.9}{\textcolor{color_pwd_2}{$\square$}}) nonbreaking solitary\WS{-like} waves, as well as for most of the forming (\textcolor{color_pwd_4}{$\blacktriangle$}) spilling breaking bores. However, the vast majority of the (\textcolor{color_pwd_5}{$\blacktriangleright$}) plunging breaking bore waves and (\scalebox{1}{\textcolor{color_pwd_6}{$\blacklozenge$}}) water jets depart significantly from the modelling. This observation is completed by the analysis of the ratio $A_{m0}/A_{\rm{Joo}}$, shown as a function of the relative acceleration $\gamma/g$ in figure \ref{wave_amplitude_end_generation_joo}(b). One may note that $A_{m0}/A_{\rm{Joo}}$ increasingly deviate from unity as $\gamma/g$ increases, which is expected due to the small relative acceleration assumption in the theory of \citet{1990_joo}. Furthermore, the most non-linear wave regimes [(\scalebox{0.9}{\textcolor{color_pwd_2}{$\blacksquare$}}) breaking solitary\WS{-like} waves, (\textcolor{color_pwd_5}{$\blacktriangleright$}) plunging breaking bores and (\scalebox{1}{\textcolor{color_pwd_6}{$\blacklozenge$}}) water jets] tend to have a maximum wave amplitude larger than the analytical prediction. If it is not obvious to define a clear criterion for the applicability of the analytical model, one can observe that the data located at $\gamma/g \lesssim 0.2$ seems to be fairly well distributed around the plateau value $A_{m0}/A_{\rm{Joo}}=1$ with a typical dispersion of order $\pm 10$ \%. Therefore, this sets an upper limit for the validity of the theory developed by \citet{1990_joo} for the present configuration. Strictly speaking, an additional assumption is made in the work of \citet{1990_joo}, which is that of a small displacement of the vertical wall. This translates into $x_p/h \ll 1$ which, at the time $t=L/U$, implies $\Lambda_p \ll 2$. However, the theory gives an accurate estimate of the wave maximum amplitude even in the case of the (\textcolor{color_pwd_4}{$\blacktriangle$}) spilling breaking bore waves, for which the value of $\Lambda_p$ is always greater than 2. We infer from this observation that the small piston displacement assumption has a minor influence \WS{on the free surface elevation} in the vicinity of the translating wall, and thus is not critical for estimating $A_{m0}$.

Therefore, the theory developed by \citet{1990_joo} efficiently describes the generated waves for small relative accelerations of the piston such that $\gamma/g \lesssim 0.2$, both in terms of the overall free surface elevation $\eta(x,t)$ and the maximum wave amplitude $A_{m0}$ at the contact with the translating wall. Nevertheless, this approach fails at larger values of $\gamma/g$, especially in cases where a hydrodynamic shock occurs, as highlighted in figures \ref{transient_bump_growth}(c) and \ref{transient_bump_growth}(d). 


\subsection{Large Froude number: the unsteady hydraulic jump}
\label{Subsec_model.2}

\begin{figure}
	\centerline{\includegraphics[width=0.5\linewidth]{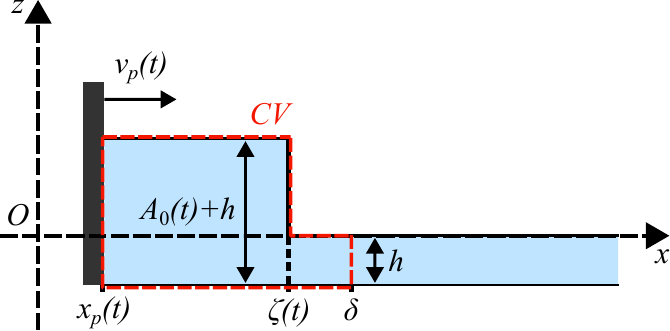}}
	\caption{Schematic of an idealized transient hydraulic jump of upstream and downstream elevations $A_0(t) + h$ and $h$, respectively. Here, $x_p(t)$ is the distance travelled by the piston at time $t$, $v_p(t)$ its horizontal velocity, $\zeta(t)$ the location of the shock and $\delta$ the position of the shock at the end of the generation phase\WS{, \textit{i.e.}, $\delta \equiv \zeta(t=L/U)$}. The letters CV stands for control volume (per unit width), which is highlighted by the red contour.}
	\label{schema_transient_jump}
\end{figure}

As already discussed beforehand, non-linear effects induce a steepening of the wave near the end of the generation phase for large values of the Froude number $\mathrm{Fr}_p$. This feature is pronounced for the plunging breaking bores and the water jets, as illustrated in figures \ref{bump_rise}(c) and \ref{bump_rise}(d). This observation is reminiscent of the formation of a hydrodynamic shock close to the source region, which is the response of the fluid to the impulse motion of the translating wall. To capture the physical mechanism at play in those circumstances, we consider the idealized case of a two-dimensional unsteady shock located at position $\zeta(t)$ at time $0 < t < L/U$, which separates an upstream region of thickness $A_0(t)+h$, for $x_p(t)<x<\zeta(t)$, from another one of thickness $h$ located downstream, \textit{i.e.}, for which $x>\zeta(t)$. Such a scenario is illustrated by the schematic shown in figure \ref{schema_transient_jump}. \WS{It will be assumed} here that (i) the flow is purely horizontal and invariant along the $z$ direction, that (ii) the fluid is inviscid and incompressible, and that (iii) the pressure can be considered as hydrostatic, \textit{i.e.}, $p(x,z,t)=\rho g [\eta(x,t) - z]$. Conditions (i) and (iii) correspond to the shallow water approximation, which implies that the vertical acceleration of the fluid is neglected \citep{1990_acheson}, and (ii) means that no dissipation occurs at the bottom of the flume during the bump formation. \WS{The fixed downstream position $\delta$, indicated in figure \ref{schema_transient_jump} and corresponding to the right side of the control volume (CV), is chosen} as the location of the shock along the $x$ axis at the end of the generation phase, \textit{i.e.}, $\delta \equiv \zeta(t=L/U)$. Therefore, the volume of fluid comprised between the abscissa $x_p(t)$ and $\delta$ and between the altitudes $-h$ (bottom of the water tank) and $\eta(x,t)$ always contains the same water particles during the generation process. \WS{From a calculation detailed in Appendix \ref{appendix_A},} the mass and horizontal momentum conservation equations applied to such an unsteady control volume read

\begin{gather}
\label{shock_mass_cons}
\Bigl( A_0(t) + h \Bigr) \Bigl( \dot{\zeta}(t) - v_p(t) \Bigr) - h \dot{\zeta}(t) = 0, \\
\notag
\\
\label{shock_momentum_cons}
\Bigl( A_0(t) + h \Bigr) \Bigl( \dot{\zeta}(t) - v_p(t) \Bigr)^2 - h \dot{\zeta}(t)^2 + \frac{1}{2} g \Bigl( \bigl( A_0(t) + h \bigr)^2 - h^2 \Bigr) = 0.
\end{gather}

\noindent The combination of these two equations leads to the following non-linear relation between the instantaneous Froude number $\mathrm{Fr}(t)=v_p(t)/\sqrt{gh}$ and relative amplitude of the bore $A_0(t)$:

\begin{equation}
\mathrm{Fr}(t) = \frac{A_{0}(t)}{h} \left(\frac{1 + A_{0}(t)/\left(2h\right)}{1 + A_{0}(t)/h}\right)^{1/2}.
\label{HJeq(t)}
\end{equation}


\noindent \WS{Equation \eqref{HJeq(t)} constitutes a transient version of the classical bore relationship \citep{1989_synolakis,1999_whitham}. \WS{Furthermore, a comparison between this expression and the prediction for a stationary hydraulic jump given by equation \eqref{HJEq} reveals the quasi-static behaviour of the unsteady hydraulic jump.} As this equation corresponds to a third-order equation for $A_0/h(t)$, the only} explicit expression for $A_0/h(t)$ \WS{as a function of Fr($t$) which is physical is therefore}

\begin{equation}
  \frac{A_0(t)}{h} =\frac{2}{3} \left\{ 2 \sqrt{ 1 + 3 \mathrm{Fr}(t) ^2 /2 } \, \cos \left( \frac{1}{3} \cos^{-1} \left[ \frac{3}{4}\frac{ \left( \mathrm{Fr}(t)-2\sqrt{2}/3 \right) \left( \mathrm{Fr}(t)+2\sqrt{2}/3 \right)}{ \left( \mathrm{Fr}(t)^2+2/3 \right) \sqrt{ 1 + 3\mathrm{Fr}(t)^2 /2 }} \right] \right) - 1 \right\}.
  \label{A0_jump}
\end{equation}

\noindent \WS{It should be emphasised that the equations \eqref{shock_mass_cons}-\eqref{A0_jump} are obtained for the generation phase, \textit{i.e.}, when $tU/L \leqslant 1$}. Interestingly, as during this stage \WS{the piston velocity is} $v_p(t)=U^2 t/L$, equations \eqref{HJeq(t)} and \eqref{A0_jump} predict that the maximum amplitude $A_{m0}$ reached at $t=L/U$ is independent of the piston stroke $L$.

\begin{figure}
	\centerline{\includegraphics[width=\linewidth]{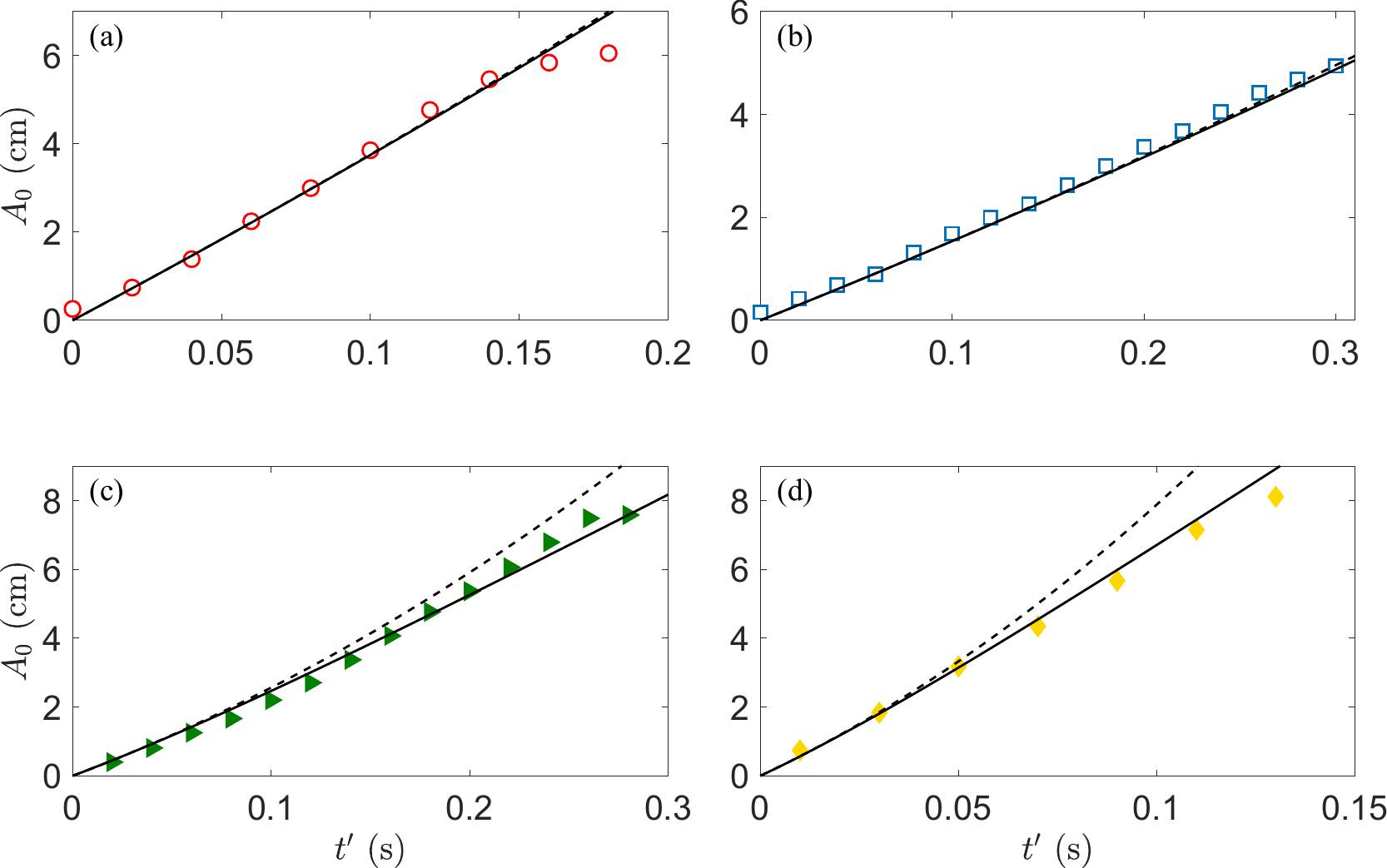}}
	\caption{Wave amplitude $A_0$ at the contact with the piston as a function of the adjusted time $t^\prime$, for the four experiments of figure \ref{transient_bump_growth}. Only the generation phase is represented here and $t^\prime$ is defined as $t-t_0$, where $t_0$ corresponds to the starting time of the generation process. The solid (--) \WS{and dashed (- - -) lines} are the predictions given by equations \eqref{A0_jump} and \eqref{PlateLawTransient}, respectively. The symbols and their colours are the same as in figures \ref{A0_t}(a)-(d).
	}
	\label{A0_t_comp_qsHJ}
\end{figure}


In order to compare the outcomes of \WS{this modelling} to the experiments, figure \ref{A0_t_comp_qsHJ} displays the temporal evolution of $A_0$ during the generation phase for the four experiments of figure \ref{transient_bump_growth}. In each plot, the solid line is the prediction given by the quasi-static hydraulic jump theory presented in this section, that is obtained by evaluating equation \eqref{A0_jump}. The time $t^\prime$ corresponds to $t-t_0$, where $t_0$ has been slightly adjusted for the experimental curves to initially coincide with the analytical ones, as the acquisition frequency of the camera ($50$ Hz) did not allowed us to determine accurately the starting time of the generation process. Despite the strong underlying assumptions of the present modelling, there is a striking agreement between the theoretical predictions and the experimental water bumps that will eventually evolve to (c) a bore wave and (d) a water jet, for which the Froude number $\mathrm{Fr}_p$ is large ($\mathrm{Fr}_p=2.0$ and $\mathrm{Fr}_p=2.2$, respectively). More surprisingly, the prediction from equation \eqref{A0_jump} also captures the time evolution of the wave amplitude $A_0$ for the two cases leading to a dispersive wave [figure \ref{A0_t_comp_qsHJ}(a)] and a solitary\WS{-like} wave [figure \ref{A0_t_comp_qsHJ}(b)], even if these regimes consist during their formation of a water bump that seems to be quite different from the idealized situation considered in figure \ref{schema_transient_jump}. This suggests that the approach followed in section \ref{Subsec_model.2} remains valid for a large range of Froude numbers. \WS{In addition, the prediction} 


\WS{
\begin{equation}
\frac{A_0(t)}{h} = \mathrm{Fr}(t) + \frac{1}{4} {\mathrm{Fr}^2(t)}
\label{PlateLawTransient}
\end{equation}
}

\noindent \WS{from \citet{1986_synolakis,1989_synolakis} is also reported in dashed line in figures \ref{A0_t_comp_qsHJ}(a)-(d). This law also exhibits good agreement with the measured values for the two low Froude number cases of figures \ref{A0_t_comp_qsHJ}(a) and \ref{A0_t_comp_qsHJ}(b), where it is almost indistinguishable from equation \eqref{A0_jump}. However, the prediction from equation \eqref{PlateLawEq} slightly overestimates $A_0$ at the end of the generation phase for experiments featuring large values for the Froude number [figures \ref{A0_t_comp_qsHJ}(c) and \ref{A0_t_comp_qsHJ}(d)].}

\begin{figure}
	\centerline{
	\includegraphics[width=\linewidth]{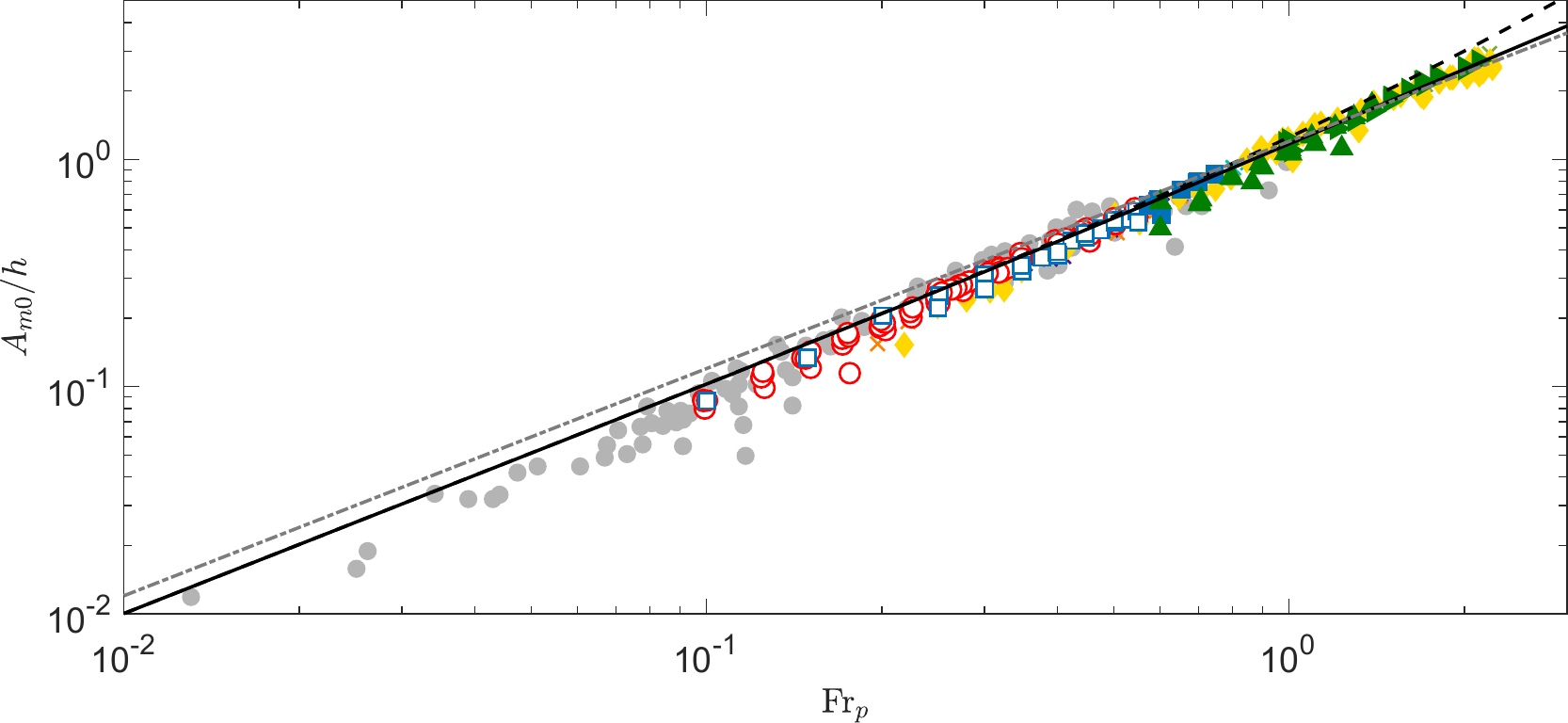}
	}
	\caption{Evolution of the relative maximum amplitude of the wave at the contact with the piston, $A_{m0}/h$, as a function of the Froude number, $\mathrm{Fr}_p$. The black solid line (\fullblack) corresponds to the quasi-static hydraulic jump prediction obtained by evaluating equation \eqref{A0_jump} at $t=L/U$, \WS{the black dashed line (\dashedblack) corresponds to equation \eqref{PlateLawTransient} from \cite{1989_synolakis}, calculated at the same time,} while the gray dash-dotted line (\textcolor{color_curve_noda}{\dasheddottedblack}) \WS{corresponds to equation \eqref{NodaEq} from} \citet{1970_noda}. \WS{All experiments from the present study are reported, with the same symbols and colours as in figure \ref{wave_regimes_diagram}(e), alongside data from \citet{1986_synolakis,1989_synolakis} (\textcolor{color_data_synolakis}{$\bullet$}).}
	}
	\label{wave_amplitude_end_generation}
\end{figure}

To further \WS{compare the experimental results with the different available models}, the relative maximum amplitude of the wave at the contact with the piston, $A_{m0}/h$, is shown as a function of the piston Froude number $\mathrm{Fr}_p$ in figure \ref{wave_amplitude_end_generation}, for all the experiments performed in the present study. Regardless of the wave regime that is obtained after the generation process, all data collapse on a master curve, highlighting once more the strong link between $A_{m0}/h$ and $\mathrm{Fr}_p$ that has already been emphasised in several past studies \citep{1970_noda,1972_das,2004_fritz,2013a_viroulet,2021a_robbe-saule,2021b_sarlin,2022a_sarlin}. As a corollary to this observation, there is no significant influence of the stroke $L$ of the piston on the value of $A_{m0}/h$. \WS{The data from \citet{1986_synolakis,1989_synolakis}, corresponding to three laws of motion for the piston (constant velocity, trajectory optimized for solitary wave generation and asymmetric parabolic trajectory), are also reported in figure \ref{wave_amplitude_end_generation} (\textcolor{color_data_synolakis}{$\bullet$}). These measurements lie on the same trend as the experiments from the present study, which suggests that the relationship between $A_{m0}/h$ and $\mathrm{Fr}_p$ is robust and independent on the detail of the forcing in a wide range of Froude numbers. Furthermore, as the studies performed by \citet{1986_synolakis,1989_synolakis} involved a much larger tank as the one used here, this confirms that there is no significant effect of capillarity in the present experiments.} The quasi-static hydraulic jump prediction, obtained by evaluating equation \eqref{A0_jump} at time $t=L/U$, is also reported in figure \ref{wave_amplitude_end_generation} in black solid line. It closely matches the experimental measurements, especially for $\mathrm{Fr}_p \gtrsim 0.4$, without any adjustable parameters. This further validates the approach of describing the generated waves as quasi-static shock waves. As equation \eqref{A0_jump} is obtained in the limit of large Froude numbers, experiments for which the shock is pronounced, \textit{i.e.}, (\textcolor{color_pwd_4}{$\blacktriangle$},\textcolor{color_pwd_5}{$\blacktriangleright$}) bore waves or (\scalebox{1}{\textcolor{color_pwd_6}{$\blacklozenge$}}) water jets, are better captured by the theoretical prediction than, for instance, ({\textcolor{color_pwd_1}{$\circ$}}) dispersive waves. \WS{The prediction given by equation \eqref{PlateLawEq} (which is equation \eqref{PlateLawTransient} from \cite{1989_synolakis} evaluated at time $t=L/U$), reported in black dashed line in figure \ref{wave_amplitude_end_generation}, is almost superimposed with equation \eqref{A0_jump} when $\mathrm{Fr}_p \lesssim 0.8$ but slightly overestimates the generated wave height for the experiments at larger values of $\mathrm{Fr}_p$.} Furthermore, equation \eqref{NodaEq} obtained by \citet{1970_noda} using the linear theory is also reported in figure \ref{wave_amplitude_end_generation} in gray dash-dotted line for comparison. Although it gives a good first order estimate of the relative maximum amplitude $A_{m0}/h$, it can be observed that the quasi-static hydraulic jump theory gives a slightly better prediction when $\mathrm{Fr}_p \lesssim 0.7$. For larger Froude numbers, the good agreement between the theory of \citet{1970_noda} and the experimental measurements is surprising as there is no proper justification for using the linear theory anymore, given that non-linear effects then become dominant.




The successful comparison between the experiments and the analytical development presented in the present section thereby suggests that, during the generation process, the observed water bumps behave as quasi-static hydraulic jumps whose vertical heights at the contact with the piston are dictated by the wall velocity $v_p(t)$ and the initial fluid depth $h$. This leads, at the end of the formation stage, to a maximum wave amplitude $A_{m0}$ which follows a weakly non-linear evolution with the Froude number $\mathrm{Fr}_p$, as highlighted by the evaluation of equation \eqref{A0_jump} at time $t=L/U$. From there, when the piston starts its deceleration, the water bump of height $A_{m0}$ relaxes into one of the different wave regimes reported in figure \ref{wave_regimes_diagram}(a)-(d).

\section{Conclusion and perspectives}
\label{sec_conclusion}

In the present study, the formation and early propagation of impulse surface waves in a water channel have been investigated experimentally at the laboratory scale. Waves were generated by the translational motion of a rigid vertical wall, which follows a constant acceleration phase followed by a constant deceleration, thereby advancing in a quadratic-in-time manner in the flume. This model experimental set-up allowed to systematically and independently vary three initial parameters: the total stroke of the piston, $L$, its maximal velocity, $U$, and the initial water depth, $h$. This was tantamount to exploring the role of two dimensionless numbers, the Froude number $\mathrm{Fr}_p=U/\sqrt{gh}$ and the relative stroke $\Lambda_p=L/h$ of the piston.

During the wave generation process, a water bump is generated in the vicinity of the source region, as a result of the translational motion of the piston, \WS{with a volume that grows with time}. This nascent wave can have a large horizontal extent and a small vertical amplitude when $\Lambda_p<1$ and $\mathrm{Fr}_p<1$ or, on the contrary, exhibits the shape of a slim \WS{and tall} water column when $\Lambda_p \gg 1$ and $\mathrm{Fr}_p \sim 1$. For a relative acceleration of the piston such that $\gamma /g=U^2/(gL) \lesssim 0.2$, the leading-order potential flow theory derived by \citet{1990_joo} gives a satisfactory description of the free surface elevation close to the advancing piston during the generation phase, alongside an accurate prediction for the maximum wave amplitude $A_{m0}$ at the contact with the rigid wall. The main observed discrepancy lies in the fact that the free surface elevation is overestimated by theory far from the translating wall, a fact that is possibly due to the assumption of a small piston displacement made by \citet{1990_joo}. Thereupon, a theoretical analysis in which this hypothesis is relaxed is needed to reach a more comprehensive description of the free-surface elevation. At larger values of $\gamma/g$, non-linear effects are no longer negligible, and the theory developed by \citet{1990_joo} then \WS{systematically} underestimates $A_{m0}$. 


For large enough Froude numbers, the non-linear steepening of the generated water bumps, especially visible at the onset of breaking, reveals the presence of a hydrodynamic shock. \WS{In that case}, a quasi-static hydraulic jump theory successfully captures the transient behaviour of $A_0(t)$ in the vicinity of the vertical wall during the whole formation stage and for all kinds of water bumps produced experimentally. As a result, it allows one to finely predict the maximum wave amplitude $A_{m0}$ reached at the end of the generation process. While this is expected for experiments where the Froude number is large, by assumption, we yet showed that the quasi-static hydraulic jump theory \WS{applies to all investigated configurations where} $\mathrm{Fr}_p \gtrsim 0.4$. This analysis shines light on the previously identified relevance of the hydraulic jump solution for describing the typical height of such perturbations \citep{1966_miller,1986_synolakis,1989_synolakis,2021b_sarlin}. As the wave amplitude $A_{m0}$ reached at the end of the formation stage can be estimated finely, it is also possible to calculate the typical horizontal extent, taken for instance as $\lambda_{m0}$ in the present study, using the fact that the dimensionless volume of the waves, $A_{m0}\lambda_{m0}/h^2$, evolves as $0.47 \, \Lambda_p$ for all experiments reported here. All these considerations open the path for a more detailed examination of the free surface elevation. In particular, it would be of great interest to investigate the spatial structure of the shock \citep{1999_whitham} more thoroughly, which could be done by relying on the quasi-static hydraulic jump model derived here. Notwithstanding the fact that this constitutes a challenging theoretical prospect, it could lead to the obtainment of the complete shape of the transient jumps during the acceleration phase of the piston.

When the translating piston begins to decelerate, the water bump of maximum amplitude $A_{m0}$ and mid-height width $\lambda_{m0}$ then relaxes into a propagating wave. In the experiments, several wave regimes are obtained and mapped in the $(\mathrm{Fr}_p,\Lambda_p)$ plane, ranging from dispersive waves obtained at small $\mathrm{Fr}_p$ and $\Lambda_p$ to unstable spilling or plunging breaking bore waves obtained at large $\mathrm{Fr}_p$ and $\Lambda_p$, which reflects the richness of the physics at play. The occurrence of stable solitary\WS{-like} waves, when $\Lambda_p$ is of order unity and $\mathrm{Fr}_p \lesssim 0.8$, denotes a situation of equilibrium between dispersion and non-linearity. More generally, in the present experiments, the value of $\Lambda_p$ selects \WS{at first order} which of these two effects will prevail, while the Froude number \Fp\ governs the relative maximum amplitude of the wave and its ``stability'' (whether it will break or not). However, when $\Lambda_p \lesssim 2.2\,{\mathrm{Fr}_p}^2$, corresponding to $\gamma/g \gtrsim 0.45$, \WS{a transition from classical wave regimes to water splashes is identified, the latter} consisting of reproducible water jets that are abruptly ejected from the vicinity of the piston. \WS{These peculiar liquid structures differ significantly from the other regimes. Preliminary} investigations suggest that these water filaments possibly undergo a ballistic flight, although this point deserves further analysis. If, in the present study, the explored accelerations of the piston were constrained by the motor limitations, it would be interesting to conduct systematic experiments at larger values of the relative acceleration $\gamma/g$ or the Froude numbers \Fp, to determine whether this conducts to the fragmentation of the water filament into large drops or not. This could help compare these water jets to other splashing phenomena as, for instance, those occurring during drop impacts \citep{2015_riboux}.

Lastly, several directions of investigation emerge from the results of the present study. On one hand, although the impulse surface wave diagram presented in figure \ref{wave_regimes_diagram}(e) extends the present state of knowledge, the region located at large $\Lambda_p$ but small Froude number was hardly accessible using the experimental setup presented here, as it resulted in very small bump amplitudes. Preliminary experiments suggest that undular bores could be observed in this situation. However, further investigations are needed to conclude on that point. A water channel and wavemaker of larger scales could be helpful to observe these waves more accurately. \WS{Moreover, to complete the present findings, a forthcoming study dedicated to the estimation of the energy budget during the wave generation and propagation processes for the different wave regimes would be of great interest, for instance using Particle Image Velocimetry (PIV) measurements.} On the other hand, studying impulse capillary waves generated, for instance, using a piston with a millimetric course, could also provide an original extension to the present work. Furthermore, it should be recalled that attention was restricted here to the case of a symmetrical forcing [see figure \ref{setup_layout}(b)]. However, symmetry breaking is expected to alter drastically the behaviour of the induced waves: we infer that this point should be further explored experimentally. In the same vein, the case of waves produced by the submarine impulsive motion of a rigid wall or by a partially-immersed piston could be of great interest, especially in the aim of comparing the results gathered to experiments involving immersed granular collapses \citep{2020_cabrera}. This could constitute a model experiment to observe the transition from shallow to deep water waves, which is of paramount interest if one keeps in mind the applications to geophysical modelling.






\backsection[Acknowledgements]{The authors are grateful to J.~Amarni, A.~Aubertin, L.~Auffray and R.~Pidoux for the elaboration of the experimental setup. They also warmly thank P.-Y.~Lagrée and M.~Rabaud for fruitful discussions.}

\backsection[Funding]{
This work has received financial support from the French Research Agency through the projet Slide2Wave ANR-23-CE30-0052-01, and from CNRS through the project TEGRAV of the interdisciplinary program MITI. 
}

\backsection[Declaration of interests]{The authors report no conflict of interest.}

\backsection[Data availability statement]{\WS{The raw data that support the findings of this study are openly available at https://doi.org/10.17882/102605, reference number 10.17882/102605.}
}

\backsection[Author ORCIDs]{

Wladimir Sarlin, https://orcid.org/0000-0002-2668-2279;

Zhaodong Niu, https://orcid.org/0009-0002-0920-693X;

Alban Sauret, https://orcid.org/0000-0001-7874-5983;

Philippe Gondret, https://orcid.org/0000-0002-7184-9429;

Cyprien Morize, https://orcid.org/0000-0002-6966-648X;
}


\appendix

\section{The quasi-static hydraulic jump}
\label{appendix_A}

%
\WS{As described in section \ref{Subsec_model.2} of the main text, the mass and horizontal momentum conservation equations applied to the control volume of figure \ref{schema_transient_jump}, under the shallow water approximation applied to an inviscid fluid, read}

\begin{equation}
\label{shock_mass_cons_0}
\frac{\mathrm{d}}{\mathrm{d}t} \int_{x_p(t)}^{\delta} \rho \left( \eta + h \right) \, \odif{x} = 0,
\end{equation}

\begin{eqnarray}
\label{shock_momentum_cons_0}
\notag
\frac{\mathrm{d}}{\mathrm{d}t} \int_{x_p(t)}^{\delta} \rho \left( \eta + h \right)u \, \odif{x} & = & \int_{-h}^{A_0(t)} p \, \odif{z} - \int_{-h}^{0} p \, \odif{z} \\
& = & \frac{1}{2} \rho g \left( \left( A_0(t) + h \right)^2 - h^2 \right),
\end{eqnarray}

\noindent where $u$ is the horizontal velocity of the fluid, which by assumption depends only on $x$ and $t$. Then, following the development made by \citet{1957_stoker} in a similar situation, one may observe that the left-hand side integrals of equations \eqref{shock_mass_cons_0} and \eqref{shock_momentum_cons_0} are of the form

\begin{equation}
\label{left-hand_integrals}
\frac{\mathrm{d}}{\mathrm{d}t} \int_{a(t)}^{b(t)} \psi(x,t) \, \odif{x} = \frac{\mathrm{d}}{\mathrm{d}t} \int_{a(t)}^{\zeta(t)} \psi(x,t) \, \odif{x} + \frac{\mathrm{d}}{\mathrm{d}t} \int_{\zeta(t)}^{b(t)} \psi(x,t) \, \odif{x},
\end{equation}

\noindent with $a$ and $b$ two continuous functions such that $a<b$ and $\psi$ corresponding either to $\rho(\eta + h)$ or to $\rho(\eta + h)u$ depending on whether equation \eqref{shock_mass_cons_0} or \eqref{shock_momentum_cons_0} is to be considered, respectively. Then, by Leibniz's rule,

\begin{eqnarray}
\label{leibniz_rule}
\notag
\frac{\mathrm{d}}{\mathrm{d}t} \int_{a(t)}^{b(t)} \psi(x,t) \, \odif{x} & = & \int_{a(t)}^{b(t)} \psi_t(x,t) \, \odif{x} + \psi \bigl( \zeta^-(t) , t \bigr) \dot{\zeta}(t) - \psi \bigl( a(t) , t \bigr) \dot{a}(t) \\
&& \mbox{} + \psi \bigl( b(t) , t \bigr) \dot{b}(t) - \psi \bigl( \zeta^+(t) , t \bigr) \dot{\zeta}(t).
\end{eqnarray}

\noindent Here, $\dot{a}(t)$ and $\dot{b}(t)$ are the horizontal components of the velocities at the contact with the vertical wall and at the end of the control volume, respectively, while $\dot{\zeta}(t)$ is the velocity of the travelling shock. Besides, $\psi \left( \zeta^-(t) , t \right)$ and $\psi \left( \zeta^+(t) , t \right)$ are the limits of $\psi$ to the left and to the right of the shock, respectively. As highlighted by \citet{1957_stoker}, in the limiting case where $a(t) \to b(t)$ but such that the discontinuity remains inside the control volume, the integral in the right-hand side of equation \eqref{leibniz_rule} vanishes. We assume that this is the case here, and apply this approach to equations \eqref{shock_mass_cons_0} and \eqref{shock_momentum_cons_0}. By considering that $a=x_p(t)$ and $b= \delta$ (so that $\dot{a}(t)=v_p(t)$ and $\dot{b}(t)=0$ as $\delta$ is constant), that $\psi \left( \zeta^-(t) , t \right)=\psi \left( x_p(t) , t \right)$ and $\psi \left( \zeta^+(t) , t \right)=\psi \left( \delta , t \right)$ and since $\eta(x=x_p(t),t)=A_0(t)$ and $\eta(x=\delta,t)=0$ by definition, one obtains

\begin{gather}
\label{shock_mass_cons_1}
\rho \Bigl( A_0(t) + h \Bigr) \Bigl( \dot{\zeta}(t) - v_p(t) \Bigr) - \rho h \dot{\zeta}(t) = 0, \\
\notag
\\
\label{shock_momentum_cons_1}
\rho \Bigl( A_0(t) + h \Bigr) \Bigl( \dot{\zeta}(t) - v_p(t) \Bigr) v_p(t) = \frac{1}{2} \rho g \Bigl( \bigl( A_0(t) + h \bigr)^2 - h^2 \Bigr).
\end{gather}

\noindent From there, by observing that 

\begin{equation}
\label{shock_relation_0}
\Bigl( A_0(t) + h \Bigr) \Bigl( \dot{\zeta}(t) - v_p(t) \Bigr) v_p(t) =\dot{\zeta}(t)^2 h - \Bigl( A_0(t) + h \Bigr) \Bigl( \dot{\zeta}(t) - v_p(t) \Bigr)^2,
\end{equation}

\noindent one eventually establishes that

\begin{gather}
\label{shock_mass_cons_2}
\Bigl( A_0(t) + h \Bigr) \Bigl( \dot{\zeta}(t) - v_p(t) \Bigr) - h \dot{\zeta}(t) = 0, \\
\notag
\\
\label{shock_momentum_cons_2}
\Bigl( A_0(t) + h \Bigr) \Bigl( \dot{\zeta}(t) - v_p(t) \Bigr)^2 - h \dot{\zeta}(t)^2 + \frac{1}{2} g \Bigl( \bigl( A_0(t) + h \bigr)^2 - h^2 \Bigr) = 0.
\end{gather}

\noindent Equations \eqref{shock_mass_cons_2} and \eqref{shock_momentum_cons_2} constitute the transient version of the classical hydraulic jump relationships, and correspond to equations \eqref{shock_mass_cons} and \eqref{shock_momentum_cons} in the main text.

\bibliographystyle{jfm}
\bibliography{bibliography}

\end{document}